\documentclass{article}

\usepackage{PRIMEarxiv}

\usepackage[utf8]{inputenc} 
\usepackage[T1]{fontenc}    
\usepackage{hyperref}
\usepackage{url}            
\usepackage{booktabs}       
\usepackage{amsfonts}       
\usepackage{nicefrac}       
\usepackage{microtype}      
\usepackage{lipsum}
\usepackage{fancyhdr}       
\usepackage{graphicx}       
\graphicspath{{media/}}     
\usepackage{natbib}
\usepackage{xcolor}

\definecolor{tempcolor}{rgb}{0.52549, 0.176471, 0.176471}

\title{Spatial verification of global precipitation forecasts
\thanks{\textit{\underline{Final peer-reviewed version published in QJRMS: \color{tempcolor}  \href{https://doi.org/10.1002/qj.5006}{https://doi.org/10.1002/qj.5006} } }}
}

\author{
  Gregor Skok \\
  University of Ljubljana, Faculty of Mathematics and Physics\\
  Jadranska Cesta 19, 1000 Ljubljana,  Slovenia \\
   \texttt{Gregor.Skok@fmf.uni-lj.si} \\
   \And
  Llorenç Lledó \\
  ECMWF \\
  Bonn, Germany \\
  \texttt{Llorenc.Lledo@ecmwf.int} \\
}

\begin{document}
\maketitle

\begin{abstract}
Verification of global high-resolution precipitation forecasts is challenging. Spatial verification techniques address some shortcomings of traditional verification. However most existing methods do not account for the non-planar geometry of a global domain, or their computational complexity is too large for global assessments. We present an adaptation of the recently developed Precipitation Attribution Distance (PAD) metric, designed for verifying precipitation, enabling its use on the Earth's spherical geometry. PAD estimates the magnitude of location errors in the forecasts employing the mathematical theory of Optimal Transport, as it provides a close upper bound for the Wasserstein distance. The method is fast and flexible with time complexity $O(n \log(n))$. Its behavior is analyzed using idealized cases and 7 years of operational global deterministic 6-hourly precipitation forecasts from the Integrated Forecasting System (IFS) of the European Centre for Medium-Range Weather Forecasts. The summary results for the whole period show how location errors in the IFS model grow steadily with increasing lead time for all analyzed regions. Moreover, by examining the time-series of the results, we can determine the trends in the score's value and identify the regions where there is a statistically significant improvement (or worsening) of the forecast performance. The results can also be analyzed separately for different intensities of precipitation. Overall, the PAD provides meaningful results for estimating location errors in global high-resolution precipitation forecasts at an affordable computational cost.
\end{abstract}

\keywords{spatial verification \and global domain \and optimal transport \and distance metrics \and displacement error}

\section{Introduction}

Forecast verification is an essential part of the developmental cycle of operational numerical weather prediction models. Despite the profound impact water cycle has on human society and the biosphere, precipitation is yet still a challenge to measure, predict and verify. Traditional non-spatial verification methods face drawbacks, such as the 'double penalty' issue, penalizing forecasts for both false alarms and missed events, while also struggling to differentiate between near misses and significant spatial displacements \citep{Brown2011,Skok2022}. Moreover, these metrics favor smooth precipitation fields that are not physically realistic but rather optimize a certain metric \citep{BenBouallgue2024}.

Spatial verification measures try to address these problems, and many such measures have been developed over the years. According to how they function they can be broadly classified into five non-exclusive categories \citep{ Gilleland2009, Dorninger2018}: neighborhood methods \citep[e.g.,][]{Roberts2008, Roberts2008a, Skok2022}, scale separation/decomposition metrics \citep[e.g.,][]{Casati2004, Mittermaier2006, Casati2010, Buschow2021, Casati2023},  feature-based approaches  \citep[e.g.,][]{Ebert2000, Davis2006a,Davis2006b, Wernli2008, Davis2009, Wernli2009}, field deformation techniques \citep[e.g.,][]{Keil2007, Keil2009, Marzban2009} and distance metrics \citep[e.g.,][]{Baddeley1992, Gilleland2017}. Most of them were originally developed when local area models reached spatial scales capable of explicitly representing convective motions. Although the exact location and timing of convective processes is difficult (if not impossible) to ascertain, it is still useful to produce forecasts that inform of the magnitude of the convection and precipitation, even if slightly out of place. 

With growing computing capacity (e.g., the advent of exascale computing with EUROHPC systems \citep{Gagliardi2019}) and recent initiatives such as Destination Earth \citep{Hoffmann2023}, global modeling is also reaching convection-resolving scales. Moreover, at longer lead times, the double penalty issue can also affect forecasts on larger spatial scales since features like cyclones and fronts can be substantially displaced in the forecasts. Thus, as the operational physics-based global models continue to be developed and improved along with the new machine-learning-based models that also show increasing potential for global forecasting \citep[e.g., ][]{Weyn2020, Bi2023,Lam2023}, there clearly exists a need for spatial metrics that could be used in the global domain. 

At the same time, to the best of our knowledge, one cannot find any examples of high-resolution global fields being analyzed by spatial metrics (that would also properly account for the spherical geometry of the Earth) in the existing literature (with maybe the exception of \citet{Mittermaier2016}). As far as we could gather, there seem to be two reasons for this; either the existing methods rely on a planar geometry and adapting them to properly account for the non-planar geometry of a global domain is difficult, or the computation complexity in spherical geometry increases considerably, making their use with current state-of-the-art global high-resolution models prohibitively expensive.

We wanted to devise a measure that would be fast enough to be used with the operational global high-resolution models, would properly account for the spherical geometry of the global domain and would not require thresholding of the input fields. We also wanted the new measure to be able to provide forecast quality information on regional and local scales, not only on the global scale. Namely, while information for the whole global domain can be of benefit (e.g., when comparing two models globally or when training a global machine-learning-based model), one is oftentimes more interested in forecast quality over specific geographic subregions (e.g., a continent, a country or a latitude belt) or surface type (e.g., land or sea). 

An intuitive way of measuring the magnitude of displacement of the predicted features compared to observations is drawing a transport plan that moves all the water volume from the forecasted locations to the observed locations. There are multiple ways to draw such a transport, but we are interested in one that requires minimal effort, i.e., the one that requires less water displacement. The mean displacement of the transport plan is known as the Wasserstein distance \citep[also known as Kantorovich-Rubinstein distance, Mallow's distance or Earth-mover distance][]{Villani2009, Rubner2000}. Computing the exact Wasserstein distance is computationally very expensive (with a time complexity of $O(n^3)$), and this explains why it has been only seldom used in weather forecasting so far \citep[some examples include ][]{Farchi2016,Nishizawa2024, francis2024examiningentropicunbalancedoptimal}. However, fast approximations of the Wasserstein distance can be computed in reasonable amounts of time. One commonly used approximation is the Sinkhorn algorithm \citep{Sinkhorn1967}, which has a time complexity $O(n^2)$, that, unfortunately, is still too large for use with current state-of-the-art operational global high-resolution models, which tend to use grids with millions of grid points in the horizontal direction. The Sinkhorn approach also has a $O(n^2)$ memory requirement, which is also prohibitively large when one needs to deal with millions of points. 

Recently, Skok presented the Precipitation Attribution Distance \citep[PAD, ][]{Skok2023}, which essentially finds out a close-to-optimal transport plan that approximates the Wasserstein distance (i.e. it does not lead to the exact distance but provides a close upper bound). In this work, we show that the PAD algorithm, which was developed and analyzed for situations where all grid points are located on a plane, can be adapted to take into account the spherical geometry of the Earth. This is achieved without a significant increase in time complexity that remains $O(n\log(n))$ (with the memory requirement $O(n)$), enabling the analysis of precipitation forecasts from current state-of-the-art operational global high-resolution models. We also show how the transport plan can be used to produce regional or local diagnostics.

\section{Data}

\subsection{IFS operational forecasts}

We analyze the quality of the deterministic (HRES) operational forecasts from the Integrated Forecasting System (IFS) of the European Centre for Medium-Range Weather Forecasts (ECMWF) produced between June 2016 and May 2023 (7 years in total) and initialized at 00 UTC. The numerical model is described in \citet{IFSchap3, IFSchap4}. Multiple IFS model versions have been employed during that period of time, but all of them share the same spatial grid. For surface fields, that is the octahedral reduced Gaussian grid O1280 \citep{Malardel2016}, which corresponds to approximately 9km of grid spacing and contains around 6.5 million grid points. Since the PAD supports irregular grids, there was no need to interpolate the fields to some regular grid, and thus, the analysis was performed using the native O1280 grid. 

All the analyses are performed on 6-hourly accumulated precipitation (expressed in $l/m^2$ or $mm$), starting either at 00, 06, 12, or 18 UTC. This model parameter includes both rain and snow and is computed as the sum of large-scale precipitation, derived from the cloud scheme, and convective precipitation, derived from the sub-grid scale convective scheme.

In order to use the precipitation attribution method for grids that do not have equal-area grid boxes it is important to first compute the total amount of precipitation (i.e., liquid equivalent volume or mass of water) in a grid box. To do so, we multiply the accumulated precipitation height by the area of the grid points. The grid box area of the O1280 is dependent on the latitude and is 61 $km^2$ at the equator, 93 $km^2$ at 75º N where it is the largest, and 18 $km^2$ close to the poles where it is the smallest (see Fig.S1 in the Supplementary Materials Section S1). 

We analyze the quality of the forecasts at four lead times, 1-, 3-, 5-, and 9-days ahead. The results for a given lead time include the four 6-hourly periods in a day. In total the analysis consists of $40\,896$ individual comparisons (7 years $\times$ $\approx$365 days $\times$ 4 time periods per day $\times$ 4 lead times).

\subsection{Pseudo-observations}

Global, high-resolution, and good-quality gridded precipitation observations are not readily available for verification despite their enormous societal importance \citep{Lledo2024}. Model analyses or reanalyses offer a convenient alternative. They provide a gap-free dataset (important for spatial verification, since most methods cannot handle those), and are less affected by measurement errors and representativeness issues than direct observations. However, analyses are a far-from perfect alternative, because they are affected by model errors. For convenience, and to illustrate the method, here we employ short-term IFS forecasts as pseudo-observations, mimicking an own-model precipitation analysis \citep{Lavers2023}. Verifying against own-analysis can lead to overly optimistic verification results because any model bias will appear in both forecast and observation fields and cancel out, but the objective of this research is not to score the IFS model but to showcase the verification method. The first 6 and 12 hours of the 00 UTC and 12 UTC HRES runs have been used to obtain 6-hourly accumulated precipitation. As with forecasts, precipitation observations are transformed to precipitation volume by multiplying them by the area of the grid box. We did not do any special treatment to avoid spin-up issues. 

Analysis of the pseudo-observations shows that for the whole globe, only 30\% of the domain is completely dry, with values ranging between 20 and 60\% depending on the region (Fig.S2a in the Supplementary Materials in Section S1). Despite this, most of the grid points in any region (>85\%) receive less than 1 mm/6h. 

One known issue of the IFS model is its tendency to produce too much light precipitation \citep{Becker2021,Lavers2021}. Therefore, areas with light precipitation will be too extensive both in the forecast and observed fields. However, an analysis of the pseudo-observations (Fig.S2b in the Supplementary Materials) shows that the total volume of water in grid points with less than 0.1 mm is negligible, and most of the precipitation volume is contributed by grid points with precipitation in the range of 1 to 10 mm/6h, therefore having a limited impact on the results we present here (except maybe in polar regions where strong precipitation is scarce).

\section{Methods}

\subsection{Precipitation Attribution Distance on a planar geometry}

The Precipitation Attribution Distance \citep[PAD, ][]{Skok2023} is a distance metric \citep[the result is provided in terms of a spatial distance, ][]{Dorninger2018} based on the random nearest-neighbor attribution concept - it works by sequentially \textit{attributing} (matching) randomly selected precipitation in one field to the closest precipitation in the other. When developing PAD, the goal was to develop a metric that would not require thresholding (but one could still use a threshold if one wished to) and would be computationally fast and numerically stable.

\begin{figure}[bt]
\centering
\includegraphics[width=9cm]{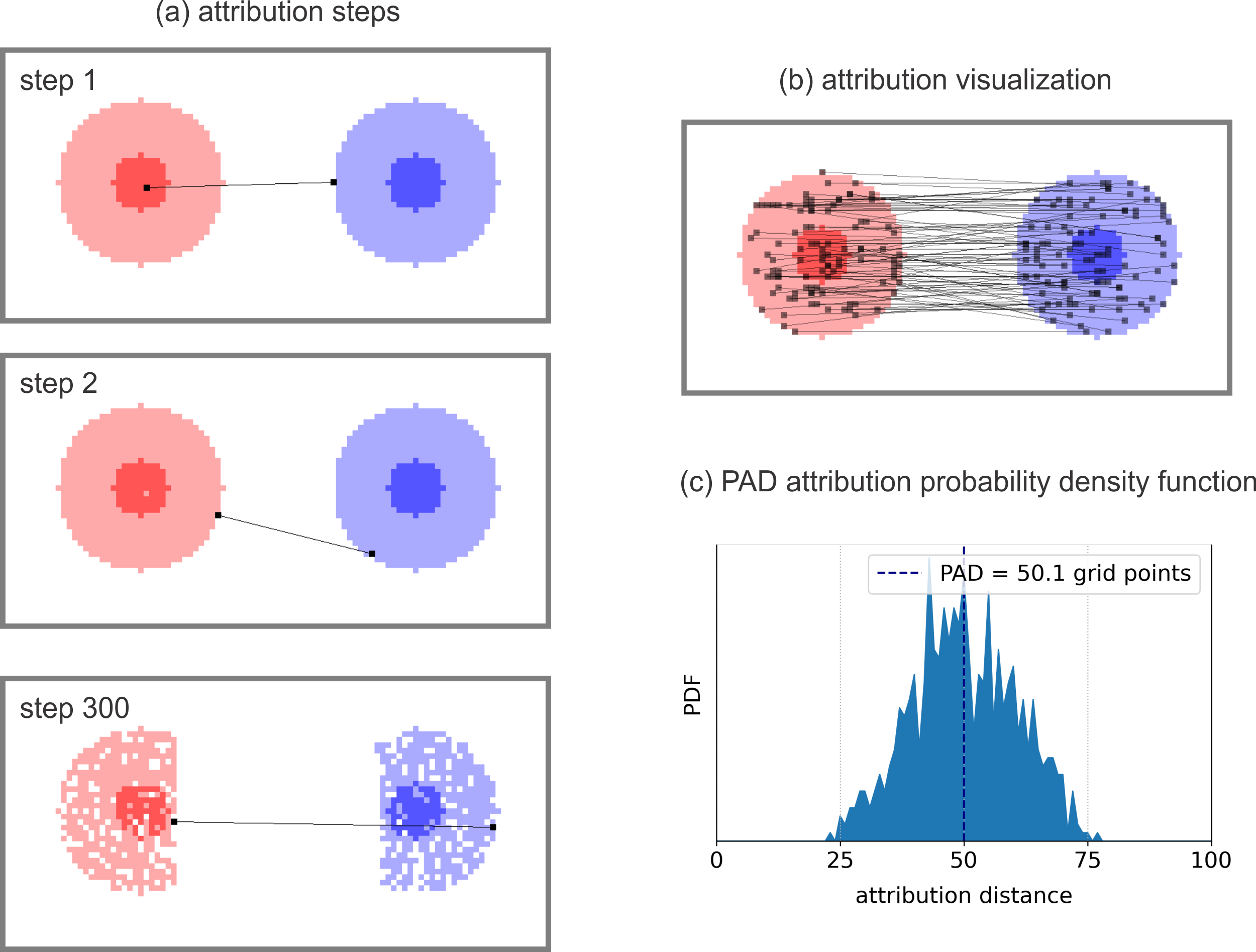}
\caption{(a) A visualization showcasing the random nearest-neighbor attribution methodology using a pair of identical double-level circular events displaced for 50 grid points. The outer region (circle with a radius of 15 grid points) has a value of 1, and the inner region (radius of 7 grid points) has a value of 2. The larger-than-zero points in the first field are highlighted in red (with the larger values shown in a darker color), while those in the second field are in blue. The first, second, and 300-th attribution steps (top, middle, and bottom, respectively) are visualized via the black line and gray squares indicating the start and end points. (b) The attribution visualization shows a subset of all attributions (only 100 randomly chosen attributions are shown) between the larger-than-zero points in two fields. (c) A probability density function of attributions with respect to the attribution distance. The mean value of the PDF (50.1 grid points), which is used to define the PAD, is denoted by the dashed vertical line.}
\label{fig:PAD_visualization}
\end{figure}

First, the two fields are normalized (e.g., by dividing each field by its sum of all values), so the sum of values in both fields is the same. This is followed by the attribution procedure, visualized in Fig.~\ref{fig:PAD_visualization}(a-b). A grid point with larger-than-zero precipitation is randomly selected in the first field (let us assume the precipitation amount at this point is $o$), and the closest larger-than-zero point in the second field is found (with precipitation amount $f$). Then the minimum of the two values, i.e., $a=\min(o,f)$, is calculated, with $a$ being the attributed amount. Also, the spatial distance between the two points $d$ is calculated (we call this the attributed distance). The values of $a$ and $d$ are saved for later use, then $a$ is then subtracted from both points, after which at least one point will have zero value (in the special case when $o=f$ both points will have zero value). 

Next, a larger-than-zero point from the second field is randomly selected, the closest larger-than-zero point in the first field is found, and a new attributed amount and distance are determined. As before, the attributed amount is subtracted from both fields. The previous two steps are repeated until all the precipitation has been attributed and the fields contain only zero points.  

Since there is a limited number of larger-than-zero points in each field, and each step removes at least one such point (by setting its value to zero), the algorithm is guaranteed to finish in $n$ steps at the latest, with $n$ being the number of larger-than-zero points in both fields. This means that there are no issues with the convergence of the algorithm, and the calculation is guaranteed to be numerically stable. 

Based on the attributed distances and using the attributed amounts as weights, a probability density function (PDF) for the attributions can be constructed (Fig~\ref{fig:PAD_visualization}c). The PAD is defined as the mean value of the attribution PDF, which is mathematically equivalent to the precipitation-weighted average attribution distance that can be expressed as

\begin{equation}
PAD =  \frac{\sum d_k a_k}{\sum a_k},
\label{eq:PAD1}
\end{equation}

where $d_k$ and $a_k$ are the attribution distance and amount for the $k$-th attribution \citep{Skok2023}. Both sums go over all the attributions. 

The computationally most expensive part of the PAD calculation are the nearest neighbor searches for the closest non-zero point in the other field that need to be performed at each attribution step. The time complexity of the so-called linear search, which computes the distances to all remaining larger-than-zero points in the other field and selects the closest one, is $O(n)$. Since approximately $n$ such searches need to be performed, the time complexity of PAD calculation via the linear search would be $O(n^2)$, which is prohibitively inefficient. Luckily, nearest neighbor searches can be sped up considerably by the use of k-d tree \citep{Bentley1975, Friedman1977, Bentley1979}. K-d tree is a multidimensional binary search tree constructed for each input field by iteratively bisecting the search space into two sub-regions, each containing half of the nonzero points of the parent region \citep{Skok2023}. The time complexity of the tree construction is $O(n\log(n))$ \citep{Friedman1977,Brown2015kdtree}. Once the tree is constructed, a nearest neighbor search for the closest non-zero point in the other field can be performed in $O(\log(n))$ expected time \citep{Friedman1977} instead of $O(n)$. Thus, the time complexity of k-d tree-based PAD calculation (including the construction of the tree and subsequent searches) is $O(n\log(n))$. Also, as the number of remaining larger-than-zero points continuously decreases with each attribution step, the subsequent searches can also be done more quickly, which helps to reduce the computation time further. 

In \cite{Skok2023}, the behavior of the PAD was analyzed in many idealized and real-world situations and was compared to the behavior of other distance metrics. The results showed that, overall, the PAD provides a good and meaningful estimate of precipitation displacement that tends to be in line with a subjective forecast evaluation. Its results can be related to the actual displacements of precipitation events, with larger or more intense events having a proportionally larger influence on the resulting value. Moreover, the attribution PDF can be used to analyze the distribution of displacements, making it possible to gain additional insights into the quality of the forecasts. For example, the uncertainty of the resulting value, the spread of the resulting distances, and the information about the portion of precipitation attributed at small or large distances. 

One beneficial property of the PAD is that the fields can be provided on irregular grids and that it can be used to directly compare two fields that are provided on different grids without the need for interpolation to a common grid. This is beneficial since the use of interpolation can, for example, smooth the fields or change the extreme values \citep{Krystyna2023}. 

Since the method randomly selects the nonzero points as they are being attributed, the behavior of the metric is nondeterministic. However, as long as the two fields contain enough points (e.g., more than 1000), the effect of the nondeterministic behavior was shown to be negligible \citep{Skok2023}. Alternatively, the PAD value can be calculated multiple times for the same two fields, and the minimum value can be used as a more robust result (in \cite{Skok2023}, we suggested that the average value could be used, but in fact, it makes more sense to use the minimum value, as this gives a better approximation of the Wasserstein distance).

The PAD produces a set of attributions that can be understood as a close-to-optimal transport map between precipitation mass in both fields being compared. Therefore, the PAD distance on a planar geometry can be understood as an upper-bound approximation of the true (optimal) Wasserstein-1 distance \citep{Villani2009, Rubner2000}. It is worth noting that exact algorithms for the computation of Wasserstein distances have time complexity of $O(n^3)$ \citep{Tomizawa1971, Jonker1987}, and this approximation is much faster.

\subsection{Computing the Precipitation Attributions on a spherical domain}

Although the possibility of using PAD in the global domain is mentioned in \cite{Skok2023}, this avenue was not explored. There are two aspects to be considered. Firstly, most spherical grids do not have equal-area grid boxes, as is the case in many limited-area model grids. In the second place, the nearest-neighbor search for the closest non-zero points in the other field needs to be adapted for the spherical geometry of the Earth, where distances are measured across great circles (along the spherically curved surface of the planet).

The first issue can be solved by attributing a physical quantity that takes into account the grid box area, such as precipitation mass or (liquid water equivalent) volume, instead of precipitation height. This is further detailed in the data section. This extension can also be useful in the case of limited area model grids on a plane but with unequal grid box areas.

The second difficulty, which is inherent to the Earth's geometry, can be easily tackled with a subtle mathematical insight that simplifies the nearest neighbor search. The key step consists of embedding all the grid points of the Earth's surface (originally a two-dimensional spherical coordinate system) into a three-dimensional Euclidean space. In this new coordinate system, computing the regular Euclidean distance between two points that lie on the Earth's surface corresponds to measuring the chord length, i.e. the straight segment that connects the two points across the Earth (sometimes referred to informally as the tunnel distance, TD). On the other hand, computing the Geodesic distance along a great circle (GCD) on the Earth's surface corresponds to the length of the arc delimited by the two points. Since a larger chord will always correspond to a larger arc and viceversa, searching the nearest neighbor with the tunnel distance yields the same result as using the great circle distance. Therefore, the nearest neighbor searches can be done using a k-d tree in the 3-dimensional space instead of using distances along the curved surface of the Earth. Once the closest point is identified, the GCD, which represents the attributed distance, can be calculated from the TD using $GCD=2 r_E \arcsin\left(TD/(2r_E)\right)$, where $r_E$ is the Earth's radius, or from the latitude/longitude coordinates of both points by using the Haversine formula \citep{Markou2010}. 

\subsection{Idealized cases on a spherical domain}

\begin{figure}
\centering
\includegraphics[width=0.9\textwidth]{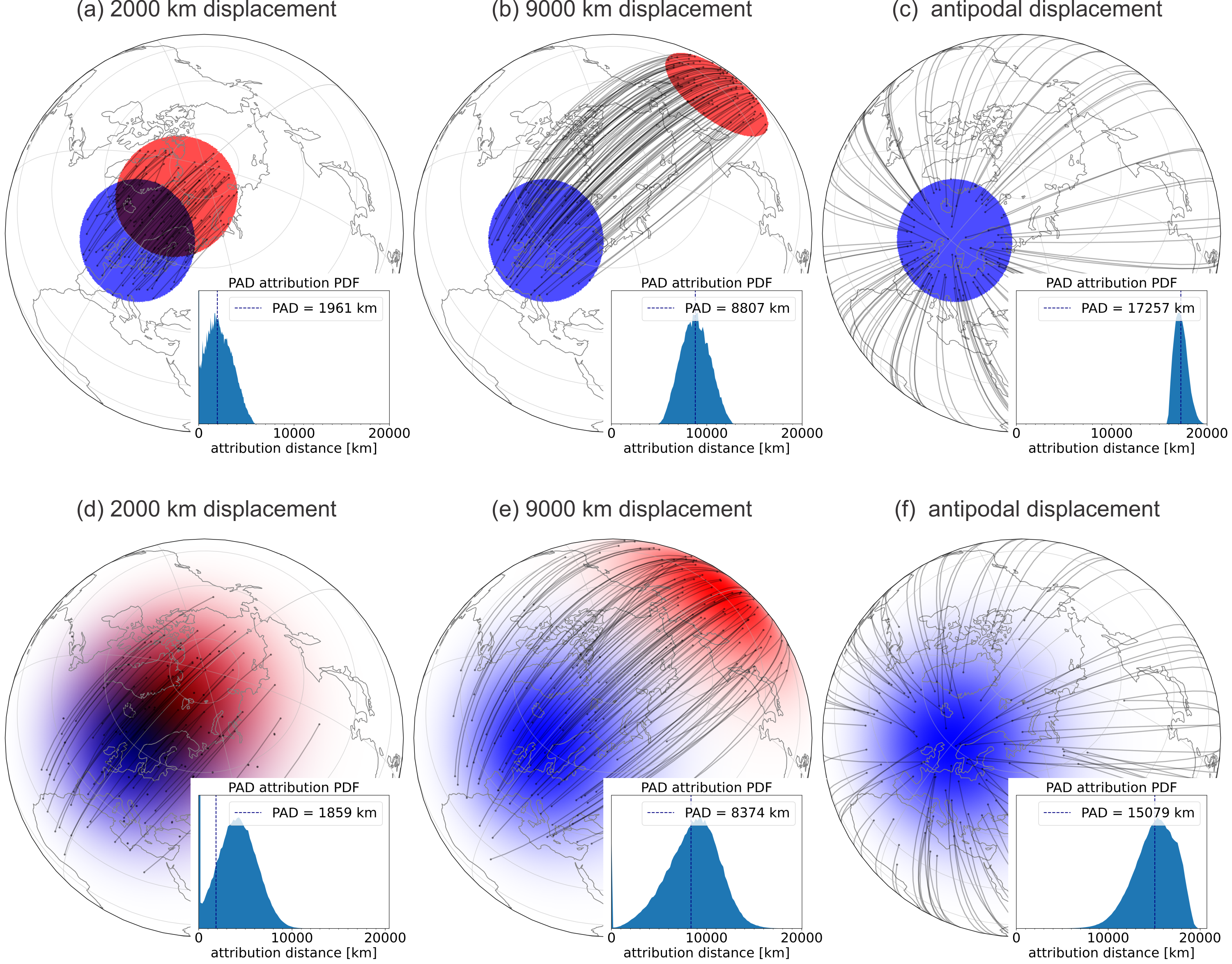}
\caption{The analysis of idealized cases with a pair of identical events with varying displacements. (a-c) Circular events with a radius of 2000 km and value 1, while the value is zero everywhere else. (d-f) Gaussian events with the sigma parameter 2000 km and amplitude of 1. The center of the blue event is always fixed at location 60N, 0E, while the red events are displaced northward with respect to the blue event. The antipodal displacement indicates a situation when the red event is located on the opposite side of the earth (with its center at 30S, 180E) - in this case, the displacement is approximately $20\,000$~km. In each sub-figure, 100 randomly chosen attributions are shown with black lines. The probability of a certain attribution being selected for visualization is proportional to its attributed amount, meaning that attributions with larger attributed amounts have more chance of being shown. All cases were constructed using a regular 0.25° latitude/longitude grid.} 
\label{fig:PAD_idealized_cases}
\end{figure}

In \cite{Skok2023}, the PAD's behavior was already thoroughly analyzed using many idealized cases, which showed that it yields a good approximation of the displacement between precipitation features. Here, we focus on the effect of spherical geometry, which represents a new aspect. Fig.~\ref{fig:PAD_idealized_cases} shows a few simple idealized cases that highlight the effect of spherical geometry - it shows the results for a pair of identical large idealized features (either circular or Gaussian in shape) with varying displacements. The displacement is either 2000 or 9000 km, or the other event is located at the antipode (on the exact opposite side of the Earth). The antipodal situation represents a limiting case since the displacement cannot be larger due to the spherical shape of the Earth. For the 2000 km displacement, the PAD values are only a bit smaller than the actual displacement between the two events, but for the 9000 km, and especially for the antipodal situation, the effect of the spherical geometry becomes more pronounced, with the PAD being noticeably smaller. However, the idealized precipitation features and their displacements in the examples shown in Fig.~\ref{fig:PAD_idealized_cases} are relatively large, and if they were smaller, the PAD value would be more similar to the actual displacement of the idealized events. 

\subsection{Normalization and unattributed precipitation}\label{sec:normalization}

The methodology presented in \cite{Skok2023} requires that pairs of fields to be compared have the same total precipitation mass/volume. That is achieved by normalizing both fields by their total sum. This ensures that all precipitation is attributed. We performed an analysis of the global bias in the IFS forecasts and found that it tends to be small - typically a few percent or less (more information is available in Supplementary Materials in Section S2 - Fig.S3). A relatively small bias is perhaps not very surprising since we compared the forecasts against pseudo-observations produced by the same model.  

However, when analyzing the IFS global precipitation fields, we noticed that the normalization could artificially introduce regional biases. For example, the forecast might initially not have a bias over Europe, but if it has a bias over the tropics, where the majority of precipitation falls, the global normalization could introduce a regional bias over Europe. This could present a potential problem since we also wanted to obtain forecast quality information on regional and local scales, not only on the global scale, and normalization could skew these results. This was the reason why we decided not to normalize the global fields when analyzing the IFS forecasts. As a result, some of the precipitation can remain unattributed (in the field that has more precipitation to start with). In the optimal transport jargon, this is known as an unbalanced optimal transport problem. 

Any forecast bias will therefore contribute to non-attributed precipitation, but this term also includes any non-systematic or regional imbalance between pairs of fields. Doing a parallel with categorical verification, the attributed and unattributed precipitation could potentially be understood in terms of hits (the attributed precipitation), misses (the non-attributed precipitation in the observed field), and false alarms (the unattributed precipitation in the forecasted field). However, categorical scores are based on the identification of distinct events, and as events are not identified within the PAD framework, one should be careful when using concepts such as hits, misses, and false alarms, to avoid wrong impressions and misunderstandings (e.g. the concept of correct rejections cannot be directly translated). 

\subsection{Attribution distance cutoff}

When analyzing the IFS forecasts, we also noticed that there could be some very distant attributions (e.g., more than $10\,000$~km). This usually happens at the late stage of the attribution process, after most of the precipitation has already been attributed, and the algorithm might need to go searching very far to find the remaining precipitation in the other field (pairing features that are physically unrelated, and hence mismatched). Usually, the amount of precipitation involved is not large, but since the PAD is the mean value of the attribution PDF, a relatively small amount of very distant attributions could disproportionately influence (increase) the mean value, thus skewing the result. 

Very distant attributions are also a conceptual problem from a meteorological perspective since it does not make much sense to, for example, attribute precipitation in a mid-latitude cyclone to precipitation in the ITCZ. If, in a forecast, a cyclone is somewhat displaced while also producing too much precipitation, it makes more sense to see this situation as a combination of a location error (displacement) and a magnitude error, as opposed to seeing it only as a location error (with part of the cyclone's precipitation being attributed to some far away precipitation system, not related to the cyclone). 

This is the reason why we use a cutoff distance (similar to the cutoff transformation used by \cite{Baddeley1992}), which allows only attributions with distances smaller than the prescribed cutoff value. The cutoff is implemented during the attribution phase; if the distance to the closest larger-than-zero point in the other field is larger than the cutoff distance, then this point is removed from the list of remaining points, and its remaining precipitation remains unattributed (there also exists an alternative less optimal post-processing-based implementation of the cutoff - see Section S3 in the Supplementary Materials for details).

The use of a cutoff distance increases the amount of unattributed precipitation and can have a large influence on the resulting PAD value. It is also the only free parameter of the PAD that needs to be specified.

\begin{figure}
\centering
\includegraphics[width=11cm]{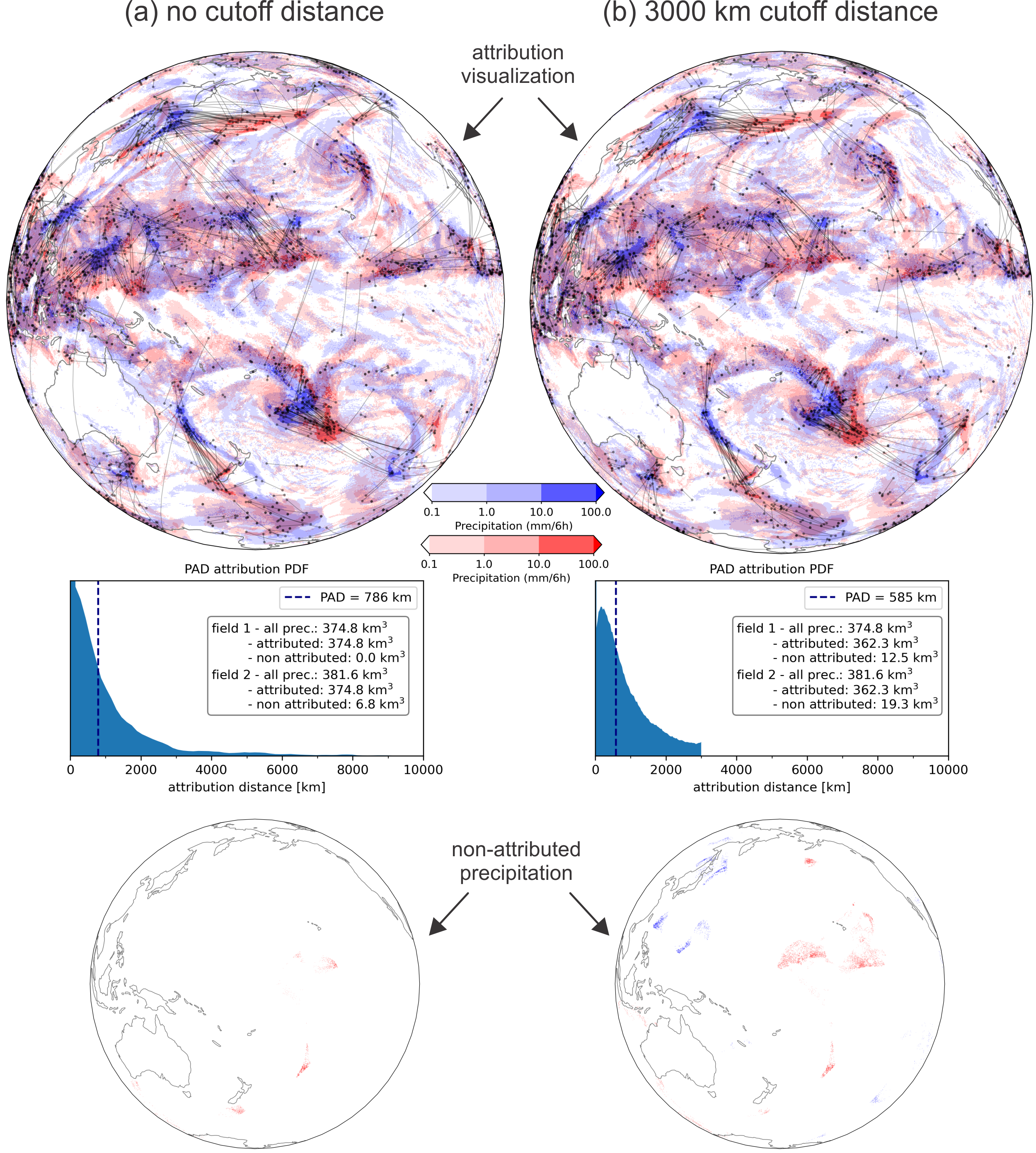}
\caption{The PAD results for the 9-day IFS forecast of 00-06 UTC precipitation for 11 October 2022. (a) shows the results where no cutoff distance for the attributions was applied, while (b) shows the results with a 3000 km cutoff distance applied. The top panel in each sub-figure shows the visualization of attributions (similarly to Fig.\ref{fig:PAD_idealized_cases} but with 2000 randomly chosen attributions shown) with the shaded blue and red colors denoting the precipitation in the pseudo-observations and forecast fields, respectively. In the middle, the PAD attribution PDF graphs are shown with an inset also showing the total, the attributed, and the non-attributed precipitation volume in each field, with fields 1 and 2 denoting the pseudo-observations and forecast, respectively. At the bottom, the spatial distribution of non-attributed precipitation is shown, with blue and red colors again denoting the precipitation in the pseudo-observation and forecast fields, respectively. } 
\label{fig:cutoff_visualization}
\end{figure}

For instance, Fig.\ref{fig:cutoff_visualization} illustrates a comparison between the results of a 9-day IFS forecast conducted with and without a cutoff of 3000 km. As can be observed, there are some very distant attributions in the variant without the cutoff, for example, between the precipitation in ITCZ and midlatitudes, which is also reflected in the attribution PDF, which has a very long tail. In the variant with the cutoff, the attributions are limited to 3000 km, with the cutoff clearly visible in the attribution PDF, and there are much fewer attributions between the tropical and midlatitude precipitation. 

The insets in the PAD attribution PDF graphs in Fig.\ref{fig:cutoff_visualization} also show the total, the attributed, and the non-attributed precipitation volume in each field, specified in km$^3$. As can be observed, the forecast overestimated the global precipitation volume by about 2\%  (374.8 vs 381.6 km$^3$). In the variant with no cutoff, the non-attributed precipitation (6.8 km$^3$) can only be found in the forecast field. In the variant with the cutoff, the non-attributed precipitation can be found in both fields (12.5 and 19.3 km$^3$) since some additional precipitation does not get attributed due to the cutoff. Nevertheless, the amount of non-attributed precipitation is still relatively small - about 5\% of the total precipitation volume. One can also notice that the red and blue colors in Fig.\ref{fig:cutoff_visualization}b bottom are never located close by. This is because all the precipitation from the two fields, which is located closer than the cutoff value of 3000 km, will always be attributed.

As expected, the resulting PAD value for the variant with the 3000 km cutoff is notably smaller than for the variant without the cutoff (585 km vs. 786 km). The larger value of the variant without the cutoff can be attributed to the long tail of the attribution PDF, which extends to very long physically-meaningless distances and is not present in the variant with the cutoff. 

Since the cutoff distance is a user-specified parameter that can substantially influence the PAD results, it is prudent that a sensitivity analysis is performed to determine to what degree the results would change if a larger or smaller cutoff value were used instead (in our case, this analysis is shown in Supplementary materials in Section S4).

\subsection{Hedging and smoothness}

Not using normalization (Section~\ref{sec:normalization}) also has a drawback, namely, it makes the method more sensitive to hedging. For example, if the fields are normalized, the PAD will be insensitive to hedging via multiplication (i.e., improving the metric's value by deliberate multiplication of all values in the field by a constant value, thereby increasing or decreasing the total amount of precipitation). To get an idea of how hedging can influence the PAD, we analyzed how two simple hedging strategies -- multiplication and addition/subtraction -- can influence the results for a single case. The multiplication only changes values where precipitation already exists, affecting higher values more. Addition or subtraction uniformly increases or decreases all values, potentially reducing zero-precipitation points in case of addition and increasing them when subtracting. The results are presented in the Supplementary Materials in Section S5.

The analysis shows that these two hedging strategies, although reducing the PAD value and indicating better forecasts, also increase the amount of non-attributed precipitation, suggesting poorer forecasts. The two strategies influence forecasts differently; multiplication benefits short-term forecasts where precipitation location is more accurate by enhancing correct values, whereas addition is slightly more beneficial for longer-term forecasts where the precipitation displacements tend to be larger will less overlap. This shows that it is important to always look not just at the PAD distance but also at the amount of non-attributed precipitation. To counteract hedging effects, several approaches are also proposed, including adjusting PAD values for forecast bias or the amount of non-attributed precipitation. 

Another aspect we tried to analyze was the influence of smoothness - for example, how the value of the metric changes if the forecasted field is smoothed. The results of this analysis are shown in the Supplementary Materials in Section S6. 

The effect of smoothness was explored through both idealized and real-world cases, and as a comparison, we performed a similar analysis using the Root-Mean-Square-Error metric \cite[RMSE, ][]{Wilks2019}. Smoothing involved using a circular kernel that conserved total precipitation but varied in diameter. In the idealized setup, with an identical but displaced precipitation event, smoothing generally increased the PAD values, especially at smaller displacements, indicating that smoothing has a greater effect on PAD when events are closer. For RMSE, the response was more complex; more smoothing at smaller displacements raised RMSE values, but at larger displacements, the impact diminished, and the RMSE sometimes became insensitive to changes, particularly with larger kernel sizes.

For real-world forecasts, smoothing showed a tendency to increase PAD values for shorter-term forecasts but had a mixed impact on forecasts at longer lead times. Smoothing appeared to have little effect on the amount of non-attributed precipitation. On the other hand, smoothing can decrease the RMSE values and make it less sensitive to forecast lead times, suggesting it could obscure differences in forecast quality between different lead times. Overall, the results show that PAD is somewhat sensitive to smoothing, and in some situations, smoothing can also decrease the PAD value somewhat. The smoothing effect on the RMSE is more problematic, however, with its value decreasing noticeably due to smoothing in some situations or becoming almost insensitive to the magnitude of displacement in others.  

\subsection{Regional and local diagnostics}

While verification in a global domain is important, providing more localized results, representing either specific regions (e.g., the tropics, Europe, or only land regions) or specific locations (e.g.,  individual grid points), can also be beneficial to understand strengths and weaknesses of forecasts in certain areas.

A simple way to obtain a PAD value representative for a specific sub-region, which we call Regional PAD (RPAD), is to first calculate the attributions for a global domain in the same way as before but then use only a subset of attributions representative for a specific region to calculate the PAD value (i.e. only the attributions for which the point in either the pseudo-observations or the forecast is located inside the selected sub-region). In this case, Equation~\ref{eq:PAD1} can be used to calculate the RPAD value, but instead of the sums going over all the attributions, the sums go over the selected subset of attributions.  

This approach also has the benefit that the attributions need to be calculated only once. Once the attribution data is available, it can be used multiple times to calculate the RPAD values for different sub-regions. 

To get a PAD value representative for a specific location (e.g., a single grid point) one can use the Localised PAD (LPAD) approach presented in \cite{Skok2023}. LPAD can only be calculated if a sufficiently long sequence of fields is available (e.g., more than 100). Similar to the RPAD, the attributions can be calculated for the whole global domain in the same way as before. Once the attribution data for all time steps in the sequence is available, the LPAD value for the grid point denoted by index $i$ can be calculated as

\begin{equation}
LPAD(i) =  \frac{\sum d_k(i) a_k(i)}{\sum a_k(i)},
\label{eq:LPAD}
\end{equation}

where $d_k(i)$ and $a_k(i)$ are the attributed distance/amount for the $k$-th attribution in which the precipitation in either the pseudo-observations or the forecast was attributed at point $i$, with the sums going over all such attributions in the entire sequence of fields. This LPAD approach differs somewhat from the one presented in  \cite{Skok2023} where only the attributions in which the precipitation in the pseudo-observations was attributed at point $i$ were used. Considering both the pseudo-observations and the forecast makes the results more robust as it avoids potentially skewed results in the presence of substantial local biases. For example, substantially more precipitation in the forecast than in the pseudo-observations could potentially result in an unrealistically small LPAD value if one would only consider the points in the pseudo-observations.

The main benefit of LPAD is that it can be calculated separately for every grid point, and the results plotted as a 2D contour map, making it possible to visually identify spatial regions where the forecast is better or worse in terms of spatial displacement.

\subsection{Useful preprocessing steps, technical implementation and benchmark}

With PAD, there is no requirement for the two fields being compared to be on the same grid. However, when this is the case, we can take advantage of it. If the points are collocated, it makes sense that the collocated precipitation volume is attributed first (for such attributions, the attribution distance is equal to zero) before the rest of the precipitation is attributed by searching for the closest non-zero points in the other field. This can be done very efficiently as a preprocessing step using a simple loop over all the points (which could also be parallelized). After this is done, the remaining precipitation in the two fields will not overlap. There are two benefits to doing this. The first is a faster computation time since fewer nearest neighbor searches for the closest non-zero points in the other field need to be performed due to the smaller number of larger-than-zero points that remain in the two fields. The second benefit is a better approximation of the Wasserstein distance value. Namely, if two points are collocated, it is always most cost-effective that the overlapping portion of the precipitation is not moved (since there is no transportation cost due to the distance being zero). We used this approach with the IFS forecasts since both fields, the pseudo-observations and the forecasts, are provided on the same grid. 

Another preprocessing step that can be done to speed up the computation is setting all precipitation values below a threshold to zero, e.g. anything below 0.1~mm/6h. As shown in Fig.S2a in the Supplementary Materials, this would substantially reduce the number of points with non-zero precipitation without substantially affecting the total precipitation volume. However, we didn't apply this technique here since computation times were already acceptable for our purposes.

A Python package that offers efficient calculation of the PAD on Sphere value and associated parameters was prepared (please refer to the Data Availability Statement for more details). The underlying code is written in C++ and includes a Python wrapper for easy use in a Python environment. A significant effort was invested to make the code as fast as possible. For example, the Bounding Box information (the minimum and maximum coordinate values of all points in the tree branch) is added to each leaf during the k-d tree construction, which makes it possible to use the so-called Bounds-Overlap-Ball test that substantially increases the speed of nearest neighbor searches for the closest non-zero points in the other field \citep{Friedman1977}.

Depending on the processor speed, it takes about 1 to 2 minutes, to calculate the attributions and the PAD value for a pair of IFS HRES fields (each consisting of $6\,599\,680$ points) on a single core. 

\begin{figure}
\centering
\includegraphics[width=6cm]{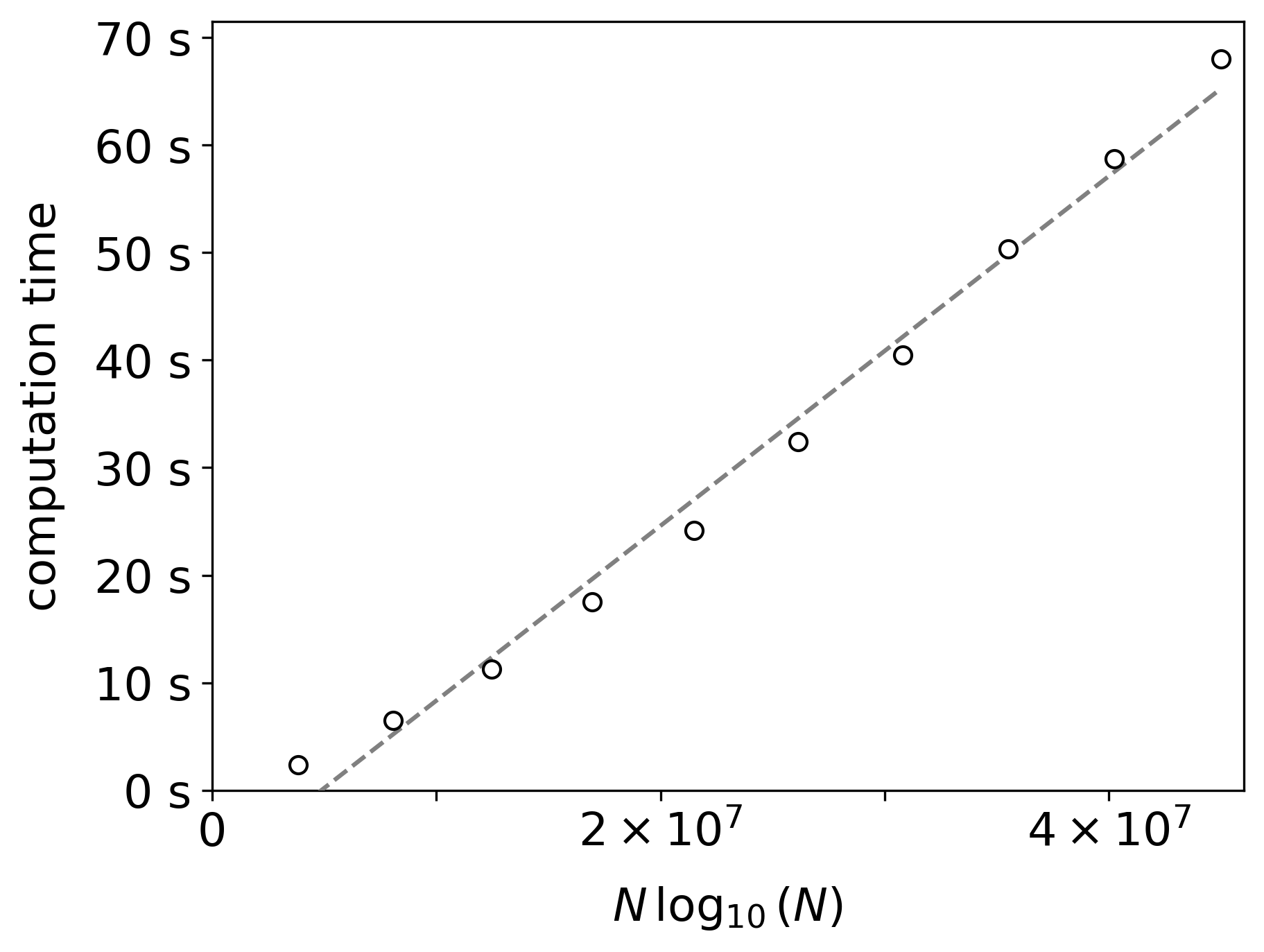}
\caption{The PAD computation time for the pair of IFS HRES fields shown in Fig.~\ref{fig:cutoff_visualization} using a 3000 km cutoff distance computed on a single core of an AMD Ryzen Threadripper PRO 5975WX processor. The computation time is shown with respect to $N \log (N)$, with $N$ being the number of points randomly sampled from the original fields that increase tenfold from 10\% ($659\,968$ points) to 100\% ($6\,599\,680$ points).} 
\label{fig:computation time}
\end{figure}

Fig.~\ref{fig:computation time} shows the PAD computation time for a pair of sub-sampled IFS HRES fields. The computation time exhibits near-linear dependence on $O(N \log (N))$, with $N$ being the number of randomly sub-sampled points, indicating that the time complexity of the approach is indeed approximately $O(n \log (n))$. 

Another aspect of computational complexity is the memory storage requirement. As explained in \cite{Skok2023}, the memory requirement for k-d tree-based PAD calculation is $O(n)$, which is also true for PAD on Sphere. 

The computational speed (and memory requirement) of PAD can be further improved via sub-sampling, at the cost of reduced accuracy. However, as shown in \cite{Skok2023}, while the sub-sampling affects the resulting PAD values, the magnitude of this effect tends to be relatively small and depends on the similarity of the two fields. For example, the resulting PAD values for sub-sampled fields used to draw Fig.\ref{fig:computation time} did not differ by more than 5 km, even though the difference in the number of points used for the PAD calculation was tenfold. 

Moreover, if multiple pairs of fields need to be analyzed, the PAD calculation can be very efficiently sped up by calculating the comparison in parallel on multiple cores simultaneously (each comparison on its own core). For example, the calculation of the 7-year analysis of the IFS HRES forecasts consists of $40\,896$ individual comparisons. It took about 3 days on a computer with an AMD Ryzen Threadripper PRO 5975WX processor to complete the PAD calculation for the entire 7-year period by parallel computation on 25 cores.

\section{Results}

Using the PAD on Sphere, we analyzed the 7-years of 6-hourly IFS precipitation forecasts, which were compared to the pseudo-observations (the corresponding 6-hourly precipitation forecast for the first day). When deciding what cutoff distance to use, besides thinking about which attributions present a conceptual problem, one can also take the typical size of features or their typical displacements into account. We used a 3000~km cutoff distance, which removes very distant attributions that do not make sense conceptually (e.g., between precipitation in ITCZ and midlatitudes) while at the same time enables detecting displacements that are smaller than the cutoff value. 3000 km is also a typical scale of synoptic features in the midlatitudes, such as cyclones and anticyclones. However, the selection of the cut-off distance is, in the end, at least partially subjective, so there does not exist a single most optimal way to define it. This is why it is important to do a sensitivity analysis to see if the main results would change if a somewhat larger or smaller cutoff were used (e.g., the ranking of results with respect to different forecast times and sub-domains, the signs of trends). In our case, a sensitivity analysis was performed using a pair of smaller and larger cutoff values (2000 and 4000~km) with the results shown in Supplementary materials in Section S4. As expected, the analysis shows that a smaller cutoff distance results in a smaller mean PAD value and more unattributed precipitation (and vice versa), but the ranking of results with respect to different forecast times and sub-domains does not change, and there are no examples in any sub-region where a trend would change its sign when a different cutoff distance is used.

First, we present an analysis of a single case valid for a specific 6-hourly period and then show how the scores over the 7-year period can be treated to assess multiple aspects of the forecast quality.

\subsection{Analysis of a single case}

\begin{figure}
\centering
\includegraphics[width=12cm]{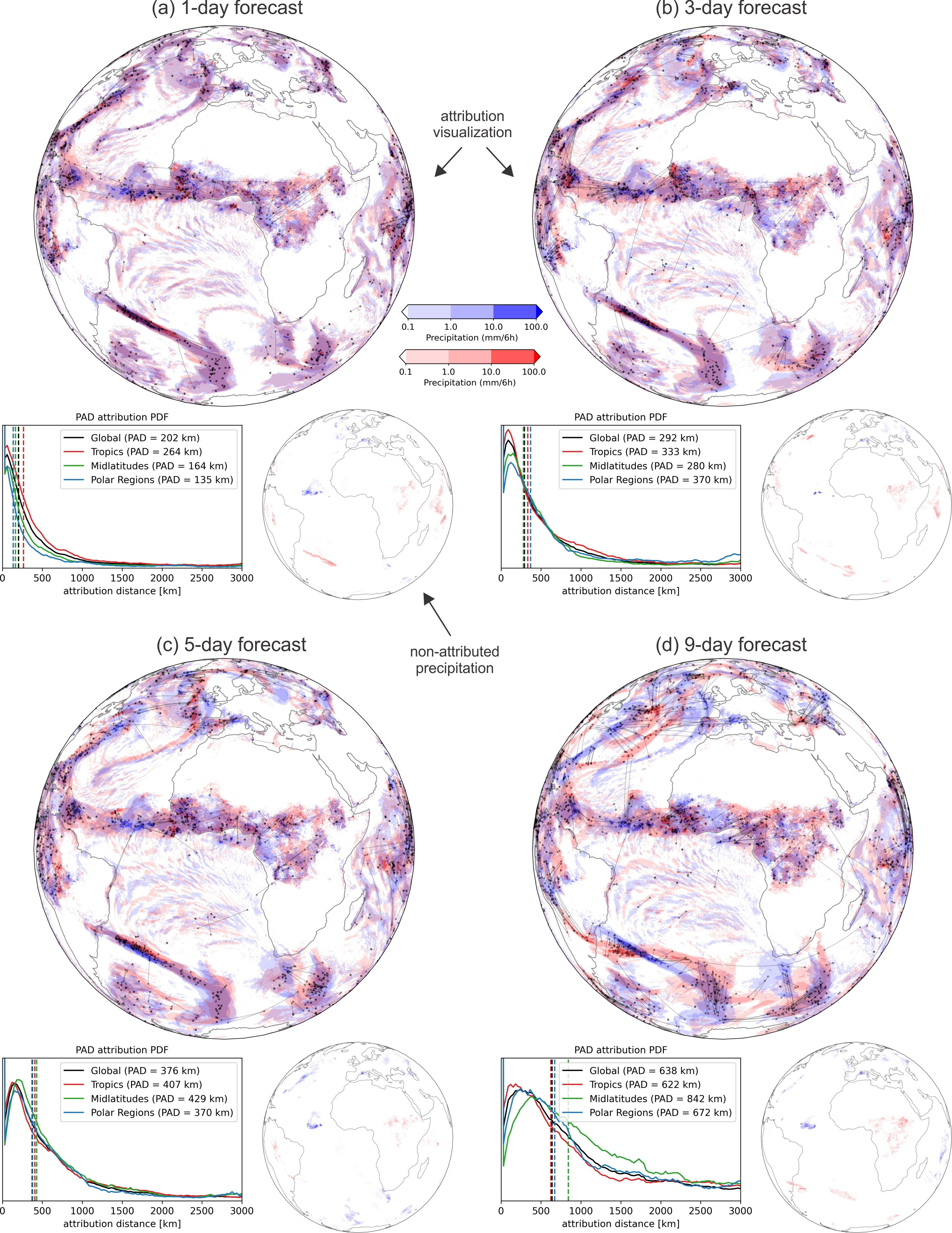}
\caption{An example of PAD results for (a) 1-day, (b) 3-day, (c) 5-day, and (d) 9-day-ahead IFS HRES forecasts of 18 to 00 UTC precipitation valid on the 21 September 2019. A cutoff distance of 3000 km was used. Similarly to Fig.\ref{fig:cutoff_visualization} the top panel in each sub-figure shows the attribution visualization, with the attribution PDF and the spatial distribution of the non-attributed precipitation shown below. In addition to the global domain, the attribution PDF and the RPAD values are shown for three sub-regions: Tropics (30S-30N), Midlatitudes (30S-60S \& 30N-60N), and Polar Regions (60S-90S \& 60N-90N). The values for the global PAD, the amount of overlap precipitation, and the amount of unattributed precipitation are shown in Table~\ref{tab:tab1}} 
\label{fig:IFS_PAD_example}
\end{figure}

Fig.\ref{fig:IFS_PAD_example} and Table~\ref{tab:tab1} show the PAD results of 1-, 3-, 5- and 9-day-ahead forecasts valid on the 21st September 2019 (6-hourly precipitation accumulation between 18 and 00 UTC). There is nothing particular about this 6-hourly period, and it was selected to showcase how the algorithm outputs can be interpreted. 

The total precipitation volume in the pseudo-observations was about 406 km$^3$/6h (Table~\ref{tab:tab1}). In contrast, the 1- and 3-day-ahead forecasts contained more, while the  5- and 9-day-ahead forecasts contained less precipitation than the  pseudo-observations. In absolute value terms the 3-day-ahead forecasts had the smallest global bias (+8.4 km$^3$/6h), while the 9-day-ahead forecasts had the largest (-23 km$^3$/6h). 

A visual comparison of Figs.\ref{fig:IFS_PAD_example}(a-d) indicates that the overlap of precipitation in the two fields decreases with lead time. The amount of overlap can be quantified with the precipitation volume attributed at zero distance, showing a decrease with increasing lead time from about 59\% for the 1-day-ahead to about 28\% for the 9-day-ahead forecasts (Table~\ref{tab:tab1}).

As one would expect, the global PAD value monotonically increases with lead time, being 202 km for 1-day-ahead and 638 km for 9-day-ahead forecasts. Notice how the PAD almost doubles from 5 to 9 days ahead, whereas the overlap (which is a non-spatially-derived quantity) only decreases by around a third. This illustrates that PAD is sensitive to the magnitude of location errors contrarily to point-to-point based methods.

\begin{table}
    \caption{PAD results for the case shown in Fig~\ref{fig:IFS_PAD_example}. The percentage values in the parentheses denote the amount of the precipitation volume (of either bias, overlapping precipitation or non-attributed precipitation) with respect to the total precipitation volume in the pseudo-observations (i.e., 406.7 km$^3$/6h).}    \centering
    \setlength\tabcolsep{3pt}
    \small
    \begin{tabular}{ccccccccc} 
    Lead time &  \multicolumn{2}{c}{\textbf{Prec. volume}} & Bias & Overlap & Global PAD & \multicolumn{3}{c}{\textbf{Non-attributed prec. volume}} \\
     & psudo.-obs. & forecast& & & & pseudo-obs. & forecast& total \\
     & [km$^3$/6h] & [km$^3$/6h]& [km$^3$/6h] & [km$^3$/6h] &  &[km$^3$/6h] &  [km$^3$/6h] & [km$^3$/6h] \\
 1 day ahead & 406.7 & 415.4 & 8.7 (2.1\%) & 238.0  (58.5\%) & 202 km & 10.3 (2.5\%) & 19.0 (4.7\%) & 29.3 (7.1\%) \\
 3 days ahead & 406.7 & 415.1 & 8.4 (2.1\%) & 199.1  (49.0\%) & 292 km & 11.6 (2.9\%) & 20.0 (4.9\%) & 31.6 (7.6\%) \\
 5 days ahead & 406.7 & 389.3 & -17.4 (-4.3\%) & 159.5  (39.2\%) & 376 km & 25.0 (6.1\%) & 7.6 (1.9\%) & 32.6 (8.4\%) \\
 9 days ahead & 406.7 & 383.7 & -23.0 (-5.7\%) & 112.6  (27.7\%) & 638 km & 38.7 (9.5\%) & 15.7 (3.9\%) & 54.4 (14.2\%) \\
    \end{tabular}
    \label{tab:tab1}
\end{table}

The amount of unattributed precipitation is another output of the algorithm that is worth analyzing. In this particular case, the total non-attributed precipitation (the sum of non-attributed precipitation from both fields) monotonically increases from about 30~km$^3$/6h for the 1-day-ahead to about 54~km$^3$/6h for the 9-day-ahead forecasts (Table~\ref{tab:tab1}). This shows an increasing amount of precipitation that cannot be assigned to less than 3000km, indicating growth of magnitude errors and/or regional biases with lead time.  The difference between the non-attributed precipitation in the pseudo-observations and the forecast reflects the positive or negative sign of the bias, with forecasts having more non-attributed precipitation compared to the pseudo-observations when the bias is positive and vice versa. 

The PAD attributions can also be used to quantify the magnitude of the location errors in specific regions. For the 1-day-ahead forecast, the displacements are largest in the tropics ($PAD = 264$ km, Fig.\ref{fig:IFS_PAD_example} ), as the model struggles to correctly forecast the location of convective precipitation in the ITCZ, while the precipitation in midlatitudes is forecasted better ($PAD = 165$ km) as the large synoptic systems, like cyclones and their fronts, are less displaced at short lead times. On the other hand, at long lead times (9-day-ahead forecast), the displacements tend to be largest in the midlatitudes ($PAD = 842$ km), where large synoptic systems can be substantially displaced in the forecast. The precipitation in the tropics is less displaced ($PAD = 622$ km) since the general position of the ITCZ, where the majority of precipitation falls, does not tend to change much in 9 days, even though individual convective events might be somewhat displaced in the forecasts.

\subsection{Aggregated statistics}

\begin{figure}[bt]
\centering
\includegraphics[width=11cm]{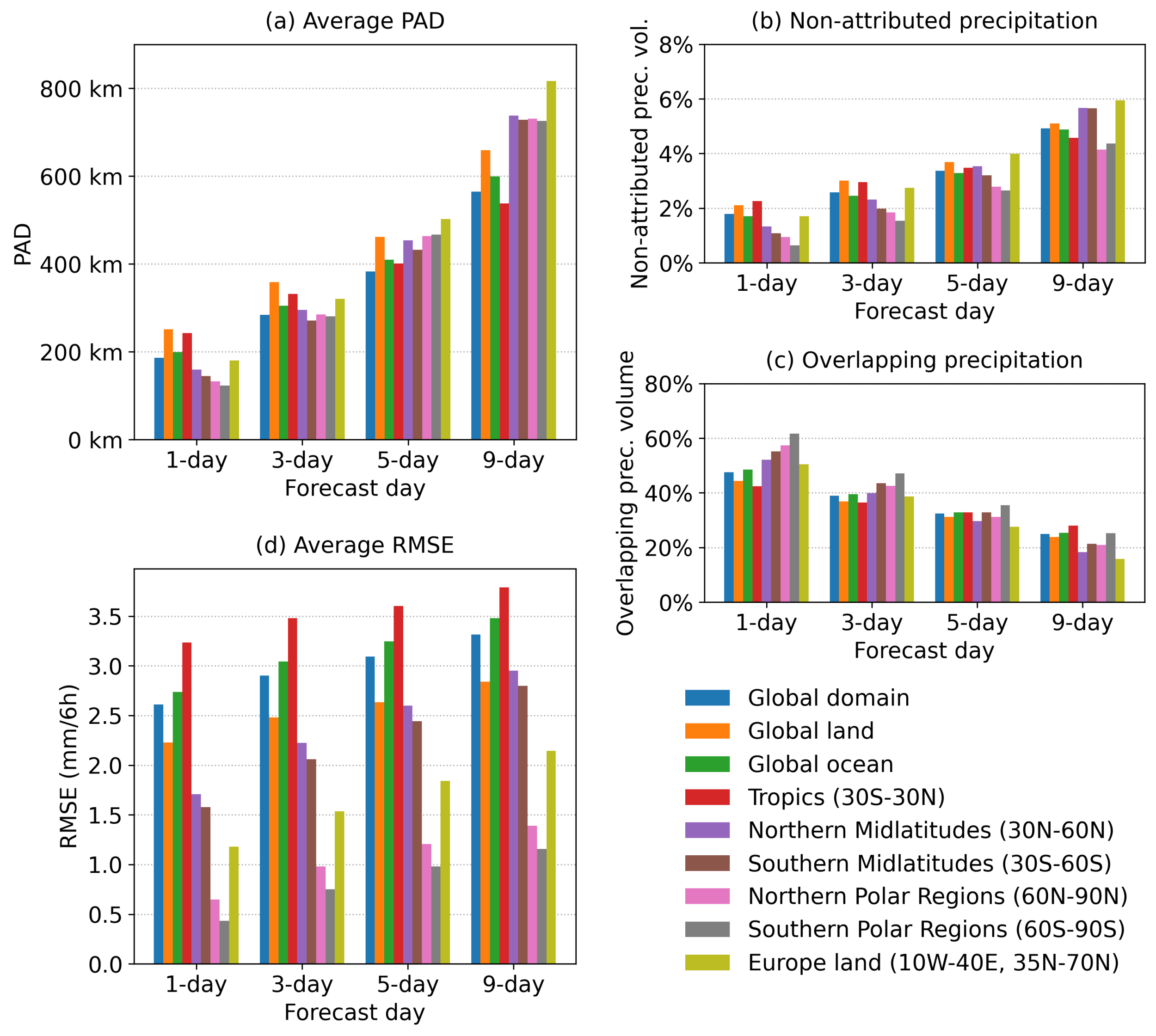}
\caption{(a) The average PAD value in the 7-year period for different sub-regions. (b) The average amount of non-attributed precipitation (the sum of non-attributed precipitation volume in both fields divided by the sum of all precipitation in both fields, expressed as a percentage) and (c) overlapping precipitation (the sum of precipitation volume attributed at zero distance in both fields divided by the sum of all precipitation in both fields, expressed as a percentage) in the 7-year period for different sub-regions. (d) Same as (a) but for RMSE.}
\label{fig:IFS_bar_graph}
\end{figure}

Fig.\ref{fig:IFS_bar_graph}a shows the average PAD value in the 7-year period for the global domain and different sub-regions. Here, the PAD values were calculated separately for each individual 6-hourly period in the 7-year period and then averaged. There is a clear distinction between the PAD values for the 1-, 3-, 5-, and 9-day-ahead forecasts. For the 1-day-ahead forecasts, the mean PAD tends to be about 180 km, depending on the sub-region, which increases to about 700 km for the 9-day-ahead forecast, meaning that the displacement errors for the 9-day-ahead forecasts are about 4 times larger than for the 1-day-ahead forecasts. With respect to the latitudinally delineated sub-regions, similar to the example shown in Fig.\ref{fig:IFS_PAD_example}, the mean PAD for the 1-day-ahead forecasts is the largest in the Tropics ($\approx230$ km), with the Midlatitues and Polar larges having a smaller value ($\approx130$ km and $\approx110$ km, respectively). For the 9-day-ahead forecast, the Midlatitudes have the largest value ($\approx675$ km), while the Tropics have the smallest ($\approx495$ km). This is in line with the analysis we described for the single case above. There is not a large difference between the hemispheres for the midlatitude and polar sub-regions. Global land shows larger displacement errors (about 15\%) than global ocean at all lead times.

The displacement error results should be complemented with information on the unattributed precipitation, which is determined by the magnitudes of global and regional biases present in the forecasts. Fig.\ref{fig:IFS_bar_graph}b shows the average amount of non-attributed precipitation in different sub-regions. Similarly to the PAD, the amount of non-attributed precipitation increases with lead time, ranging from about 2\% for the 1-day-ahead forecasts to about 6\% for the 9-day-ahead forecasts. The global ocean sub-region has less unattributed precipitation than the global land sub-region for all lead times. Moreover, the results shows that the southern hemisphere Midlatitude and Polar sub-regions have less unattributed precipitation than those of the northern hemisphere. 

The average amount of overlapping precipitation (precipitation attributed at zero distance) can also be seen in Fig.\ref{fig:IFS_bar_graph}c for different sub-regions. For the 1-day-ahead forecasts, about 50\% of precipitation is overlapping, which decreases to about 25\% for the 9-day forecasts. There is some variation with respect to the sub-regions, with the Tropics having the least overlapping precipitation of all the sub-regions for the 1-day forecasts and the most for the 9-day forecasts.

As a comparison, we performed a similar analysis using the RMSE metric. Fig.\ref{fig:IFS_bar_graph}d shows the average RMSE value in the 7-year period for the global domain and different sub-regions (same as for PAD, the RMSE values were calculated separately for each 6-hourly period and then averaged). Compared to the PAD results shown in Fig.\ref{fig:IFS_bar_graph}a, the relative differences between the RMSE values for different lead times are much smaller, while the differences between the results for different sub-regions are much larger. The large differences between different sub-regions are not surprising, since the RMSE tends to be larger in situations with more overall precipitation, as well in cases where there is more spatial and temporal variability in the precipitation field. The Tropics, where the precipitation intensities tend to be the highest compared to other sub-regions (Fig.S2 in the Supplementary materials) and variability is high, perform the worst for all the lead times, while the Southern Polar Regions, where the precipitation intensities tend to be the lowest, perform the best.

\subsection{Evolution of scores over time}

\begin{figure}[bt]
\centering
\includegraphics[width=0.95\textwidth]{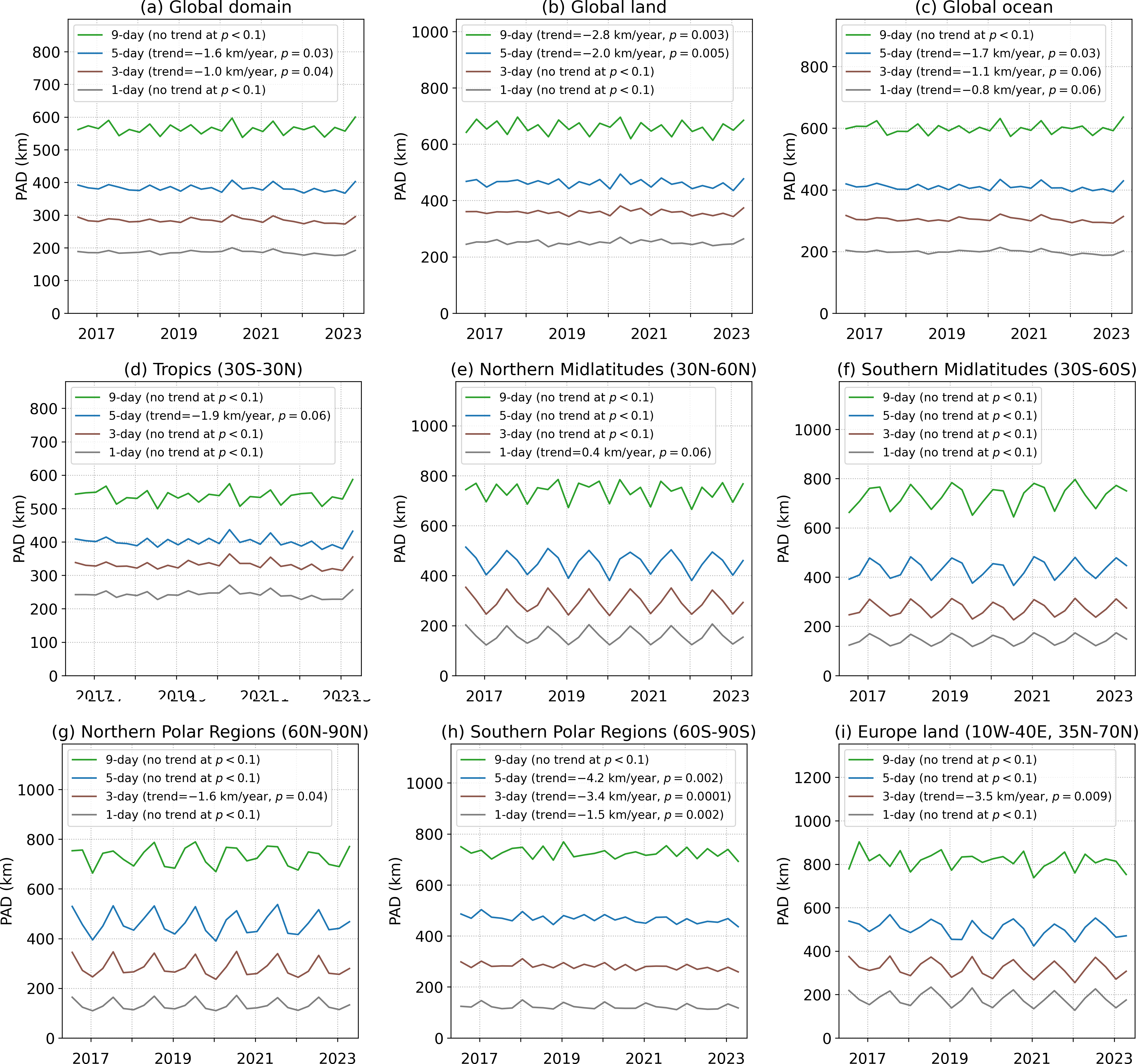}
\caption{Timeseries of mean seasonal PAD values for different sub-regions. Statistically significant trends are indicated in the legend.}
\label{fig:IFS_timeseries_seasonal}
\end{figure}

Another interesting way to look at the seven years of results is by plotting a time evolution of the scores. Fig.\ref{fig:IFS_timeseries_seasonal} shows the timeseries of mean seasonal PAD values for different sub-regions. Here, the PAD values were calculated separately for each individual 6-hourly period in a specific season and then averaged. For the extratropical regions, there is a strong seasonal cycle, showing higher displacements for the summer than for winter, coinciding with an increased convective activity. Beyond that, we can observe slight trends in the scores in some regions and lead times. A seasonal Mann-Kendall test \citep{Hirsch1982} was used to determine the statistical significance of a trend \citep[the PyMannKendall Python package was used to calculate the test and trends, ][]{Hussain2019}. A significance level of 0.1 has been used for all tests. There is a statistically significant improvement in the PAD for the 3- and 5-day-ahead forecasts for the global domain, while there is not enough evidence of a trend for the 1-and 9-day-ahead forecasts. For the global land sub-region, there is an improvement for the 5- and 9-day-ahead forecasts and no significant trends for the 1- and 3-day-ahead forecasts. For the global ocean sub-region, there is a statistically significant improvement for all lead times, except at 9 days. 

There are no statistically significant trends for the Tropics except for the 5-day-ahead forecasts that exhibit an improvement. There are no significant trends for the Midlatitudes, except a positive trend for the 1-day-ahead forecasts in the Northern Midlatitudes. In the Northern Polar sub-region, there are no significant trends, with the exception of the 3-day-ahead forecasts, which exhibit a negative trend. The Southern Polar sub-region exhibits a statistically significant improvement for all lead times, except 9 days, where there is no significant trend. The Europe-land sub-region exhibits no significant trends, with the exception of a decreasing trend (improvement) for the 3-day-ahead forecast. 

Overall, all the analyzed sub-regions exhibit either no statistically significant trends or trends indicating a slight improvement in forecast performance, with a single exception being the 1-day-ahead forecasts in Northern Midlatitudes. 

\begin{figure}[btp]
\centering
\includegraphics[width=0.93\textwidth]{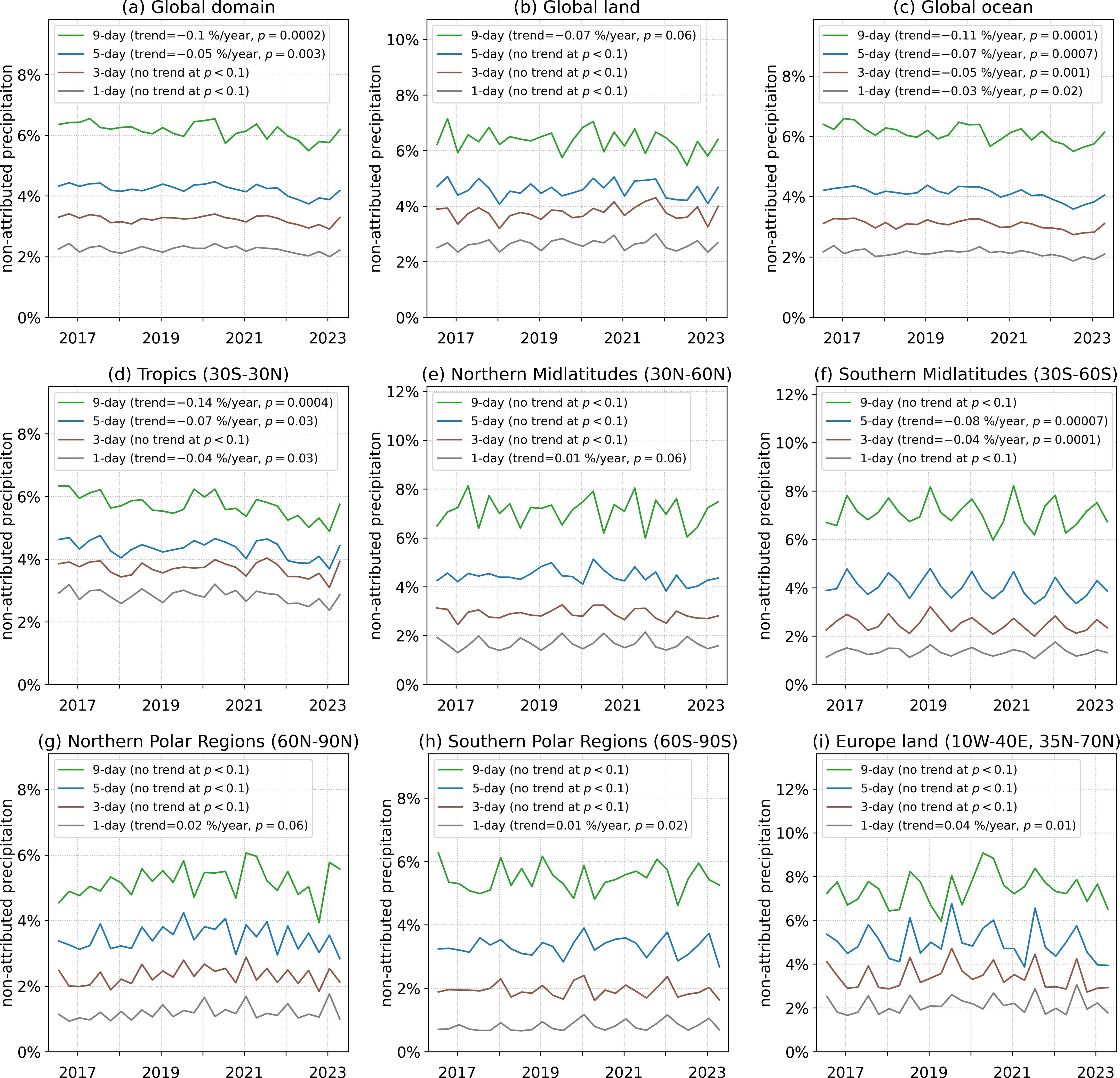}
\caption{Time series of mean seasonal non-attributed precipitation fraction (the sum of non-attributed precipitation volume in both fields divided by the sum of all precipitation in both fields, expressed as a percentage) for the 7-year period.}
\label{fig:IFS_timeseries_seasonal_NAP}
\end{figure}

In addition to the trends of PAD values, it is also important to look at the trends of non-attributed precipitation. Namely, even if the PAD would exhibit a decreasing trend, an increase in non-attributed precipitation would indicate a possibility that the forecast is not really improving overall. Fig.\ref{fig:IFS_timeseries_seasonal_NAP} shows the time series of mean seasonal non-attributed precipitation fraction for the 7-year period. As can be observed, most timeseries either do not exhibit statistically significant trends or the trends are negative. There are only four exceptions (1-day forecasts for the Northern Midlatitudes, Northern \& Southern Polar, and Europe-land sub-regions) that tend to exhibit relatively weak increasing trends of non-attributed precipitation. Based on these results, one can be more confident in the PAD-based results in terms of forecast improvement. 

Figs.\ref{fig:IFS_timeseries_seasonal} and \ref{fig:IFS_timeseries_seasonal_NAP} illustrate how the PAD can be effectively utilized to gain some knowledge on forecast improvement with time. Using data from a relatively short period of 7 years, the PAD was able to provide statistically significant information regarding the improvement (or deterioration) of forecast performance with time. As a comparison, we performed a similar analysis using RMSE (the results are in the Supplementary Materials in Section S7). In the case of RMSE, the relative differences between the score's values for different forecast times are much smaller, while the differences between the results for different sub-regions are much larger. Also, there are almost no statistically significant improvements in forecast performance in terms of RMSE, with many sub-regions exhibiting deteriorating forecast performance for most or all forecast times.

\subsection{Analysis of errors by intensity}

\begin{figure}[bt]
\centering
\includegraphics[width=0.95\textwidth]{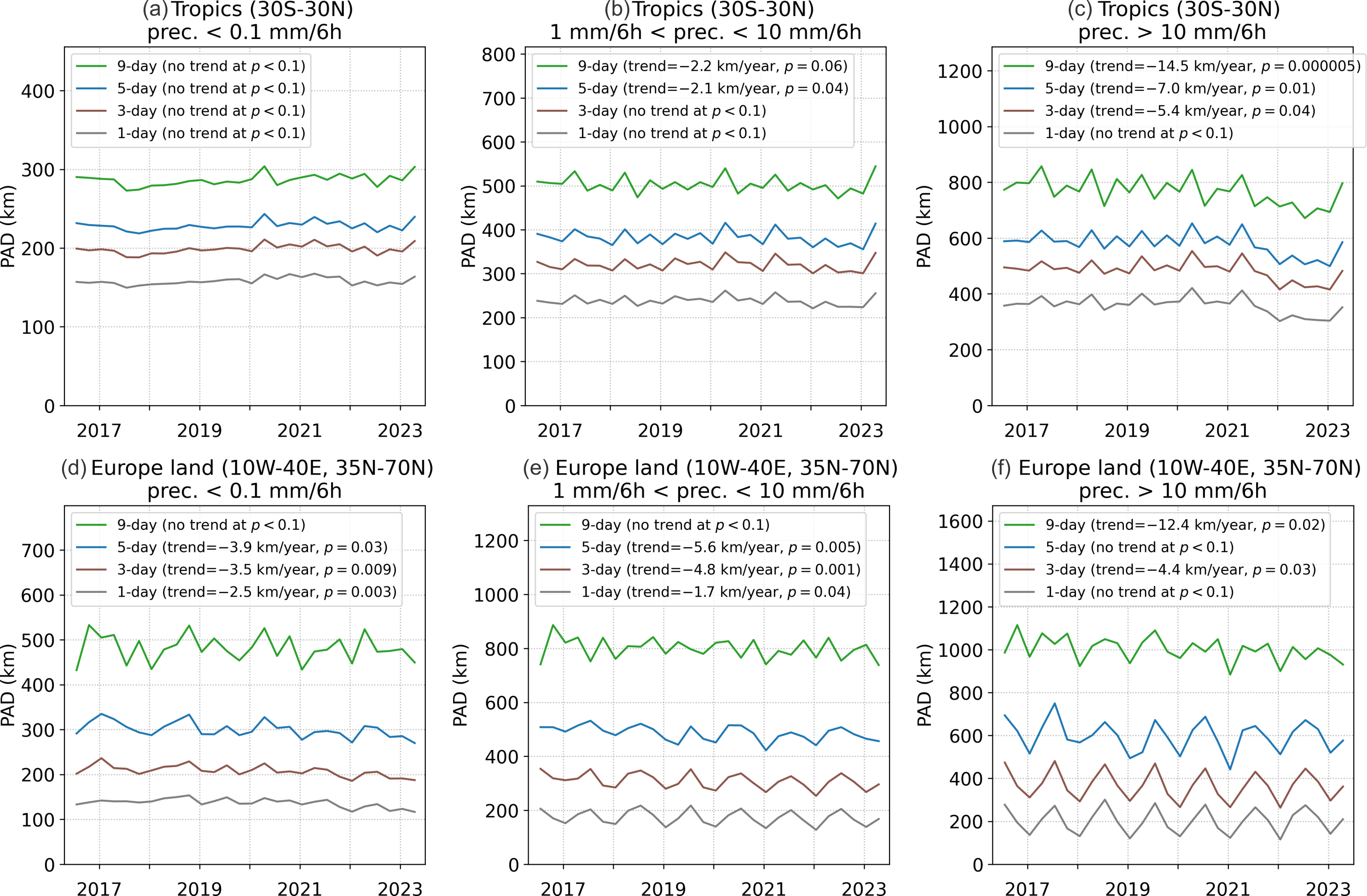}
\caption{Same as Fig.\ref{fig:IFS_timeseries_seasonal}(d,i), but with the PAD analysis done for separately for (a) low, (b) medium, and (c) high-intensity precipitation.}
\label{fig:IFS_timeseries_seasonal_intensities}
\end{figure}

The PAD attributions can also be analyzed separately for different intensities of precipitation. Fig.\ref{fig:IFS_timeseries_seasonal_intensities} shows an example for the Tropics and Europe-land sub-regions. In this case,  instead of all the attributions linked to this sub-region being used to calculate the PAD value, only attributions with the precipitation intensity (at the corresponding grid point in either the pseudo observations or the forecast) within a specific interval are used. Two thresholds (1 and 10 mm/6h) are used to separate the precipitation into three intensity ranges: low (prec. $\leq 1$ mm/6h), medium ($1$ mm/6h $<$  prec. $\leq 10$ mm/6h), and high intensity (prec. $> 10$ mm/6h). According to Fig.S2b in the Supplementary Materials, about 13\%, 56\%, and 31\% of total precipitation volume in the Tropics is contributed by the low, medium, and high-intensity precipitation, respectively. For the Europe-land sub-region these percentages are about 16\%, 70\%, and 14\%.

In Fig.\ref{fig:IFS_timeseries_seasonal}d, which shows the PAD trends for the Tropics, where all precipitation intensities are considered together, the only statistically significant trend is for the 3-day forecasts. When the analysis is done separately for different precipitation intensities (Fig.\ref{fig:IFS_timeseries_seasonal_intensities}(a-c)), there are no significant trends for the low-intensity precipitation. Still, for the medium and high intensities, there are some statistically significant trends indicating an improvement even for the longer forecast times (i.e., 5- and 9-day forecasts for the medium and high-intensity precipitation). The improvement of the high-intensity precipitation forecasts in the Tropics can be clearly seen from 2021 onwards, with reductions of up to 15 km per year.

For the Europe-land sub-region (Fig.\ref{fig:IFS_timeseries_seasonal_intensities}(d-e)), there are decreasing PAD trends for the 1-, 3-, and 5-day forecasts for the low and medium-intensity precipitation, as well as for the 3-, and 9-day forecast for the high-intensity precipitation. The results for Europe also show clearly how high-intensity precipitation (related to convection) has a stronger annual cycle than low-intensity. The results in Fig.\ref{fig:IFS_timeseries_seasonal_intensities} highlight how additional insights about the forecast performance can be obtained by analyzing separately the precipitation of different intensities. 

\subsection{Local diagnostics}

\begin{figure}[btp]
\centering
\includegraphics[width=0.95\textwidth]{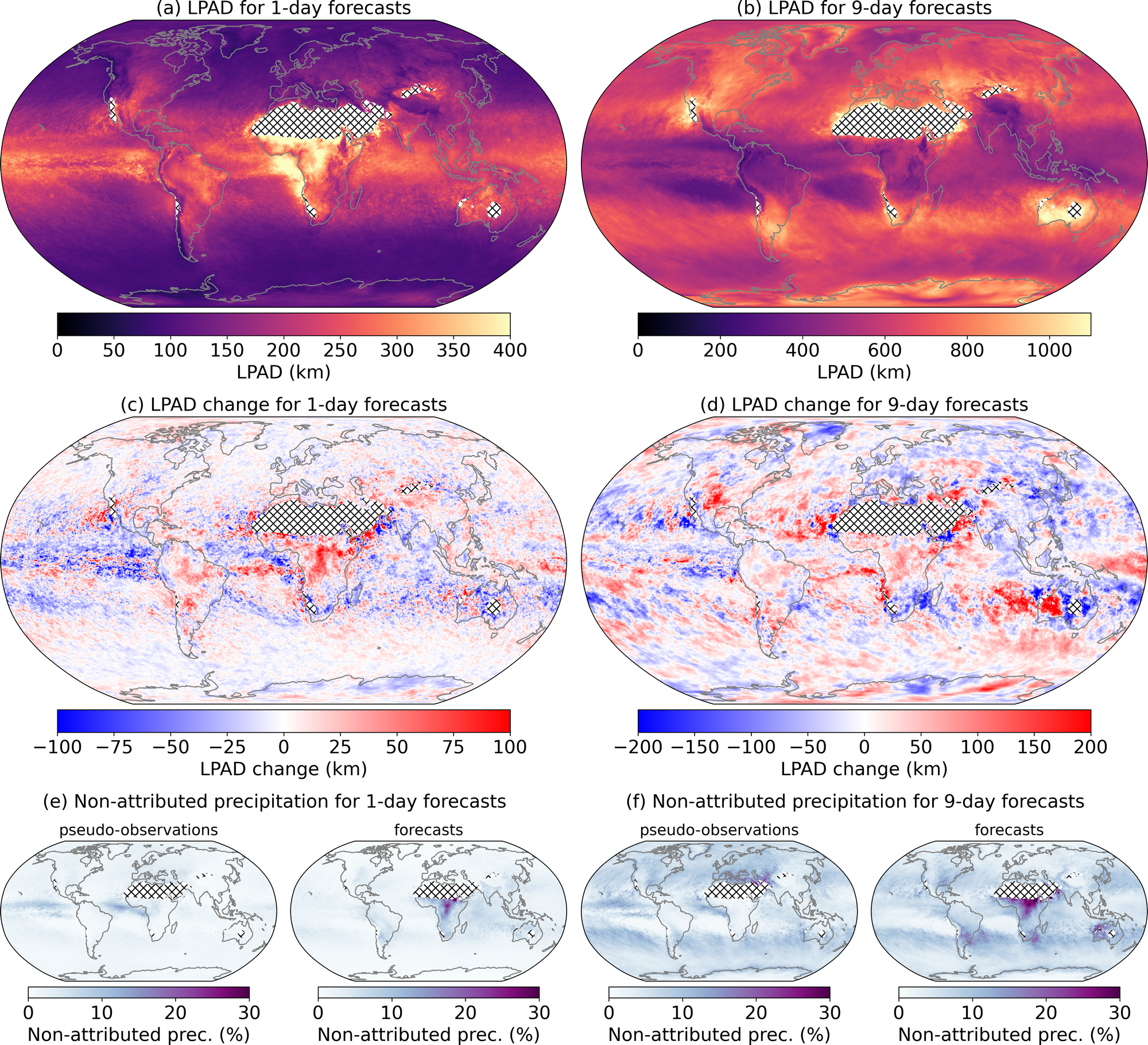}
\caption{The LPAD results for the 1-day-ahead (left) and 9-day-ahead (right) forecasts. The top panels (a,b) show the LPAD values for the 7-year period, while the middle (c,d) show the change in LPAD values between the first 4 years and the last 3 years (the blue colors represent a decrease in LPAD values, indicating an improvement of the forecast performance with time). For visualization purposes, the LPAD values were interpolated from the IFS HRES native grid to a regular $0.1^\circ$ latitude-longitude grid and smoothed using a $0.5^\circ$ square-shaped averaging kernel. The cross-hatched pattern overlays the regions where precipitation was falling in less than 10\% of the timesteps. These tend to be dry desert regions where precipitation falls only rarely, which makes the LPAD results unreliable. The bottom panels (e,f) show the fraction of non-attributed precipitation for the pseudo-observations and the forecasts for all locations in the 7-year period (the sum of non-attributed precipitation volume at a particular location divided by the sum of all precipitation at the same location, expressed as a percentage).}
\label{fig:IFS_LPAD_global}
\end{figure}

Statistics of location errors can also be compiled at grid point scale if a long enough sample is available. Fig.\ref{fig:IFS_LPAD_global}(a-d) shows the LPAD results for the 1- and 9-day-ahead forecasts (results for the 3- and 5-day-ahead forecasts are shown in the Supplementary Materials in Section S8 - Fig.S13). The top panels show the LPAD values. There is a large spatial variability of the LPAD values and a large difference between the 1- and 9-day-ahead forecasts. For the 1-day-ahead forecasts (Fig.\ref{fig:IFS_LPAD_global}a), the largest LPAD values tend to be in the Tropics, especially at or near the ITCZ, while elsewhere the LPAD values tend to be smaller. For the 9-day-ahead forecasts (Fig.\ref{fig:IFS_LPAD_global}b) the situation is reversed, with the largest LPAD values in the Midlatitudes and Polar regions. This is in line with the RPAD analysis shown in Fig.\ref{fig:IFS_bar_graph}.

The middle panels in Fig.\ref{fig:IFS_LPAD_global} show the change in LPAD values between the first 4 years and the last 3 years. The negative values (blue color) indicate an improvement, and the positive (red) indicate a deterioration of the forecast performance. For the 1-day-ahead forecasts (Fig.\ref{fig:IFS_LPAD_global}c), the most dominant red region is in central Africa, indicating that the forecast performance decreased there. In contrast, the forecast in the tropics improved over most of the oceans. There are also other extratropical regions where the forecast improved (e.g., parts of North America, Europe, South Africa, and South Asia). 

The results for 9-day-ahead forecasts (Fig.\ref{fig:IFS_LPAD_global}d) are different from the 1-day-ahead forecasts. There is not one single dominant region that stands out in terms of a deterioration of the forecast performance, but instead, there are multiple such regions located at various places around the globe (e.g., Western Australia with a nearby section of the Indian Ocean, Atlantic Ocean at approximately 30N, the central part of the USA). On the other hand, the forecast has markedly improved for Southeast Asia, the eastern part of Australia, South Africa, and parts of Europe and the Maritime Continent.

The bottom panels in Fig.\ref{fig:IFS_LPAD_global} show the fraction of non-attributed precipitation for the pseudo-observations and the forecasts. For the 1-day-ahead forecasts (Fig.\ref{fig:IFS_LPAD_global}e), the fraction of non-attributed precipitation tends to be below 5\%, although there are some exceptions. In the pseudo-observations, the largest fraction, with values up to 15\%, can be found over some oceanic regions near the ITCZ (e.g., in the Atlantic and Pacific), indicating that the forecasts underestimate precipitation there. On the other hand, the largest fraction of non-attributed precipitation in the forecasts is over central Africa, with values in excess of 20\%, indicating that the model is over-forecasting the precipitation there. 

For the 9-day-ahead forecasts (Fig.\ref{fig:IFS_LPAD_global}f) the fraction of non-attributed precipitation tends to be larger than for the 1-day-ahead forecasts. In the pseudo-observations, the regions with the largest fraction are mostly located over the oceans in the Tropics and Midlatitudes, including the Mediterranean and Southern Europe. On the other hand, the largest fraction of non-attributed precipitation in the forecasts is over central Africa, although some regions in the Midlatitudes also exhibit values larger than 15\% (e.g., the Eastern part of the North America and parts of Southern Hemisphere mid-latitudes). 

We also visualized and analyzed the LPAD results in more detail for two selected subregions - Europe and the Maritime Continent (the analysis is shown in the Supplementary Materials in Section S8). The analysis highlights how the LPAD result vary for different parts of Europe and the Maritime Continent, and are affected by geographic features such as land, sea and orographic barriers on a smaller spatial scale.

\section{Discussion and Conclusions}

Traditional point-to-point verification metrics have difficulty identifying improvements in precipitation forecasts when these are subject to location errors. On the one hand, forecasts without features can score better than forecasts with a correct but displaced feature. On the other hand, point-to-point measures cannot detect improvement in the location error magnitude itself. Optimal transport, and in particular Wasserstein distances, can be employed to overcome these two problems. However, computing exact Wasserstein distances is computationally infeasible on high-resolution global grids \citep[$O(n^3)$,][]{Tomizawa1971}. \citet{Skok2023} presented a fast metric (PAD) based on nearest-neighbor searches for the closest non-zero points in the other field that approximates the Wasserstein distance on a planar geometry. \citet{Skok2023} showed that, overall, the PAD provides a good and meaningful estimate of precipitation displacement that tends to be in line with a subjective forecast evaluation. It does not require thresholding, is not overly sensitive to noise, and its results can be related to the actual displacements of precipitation events. In this work, we have shown how the PAD can be employed on the spherical geometry of the Earth.

The method is fast and flexible. We have shown that the PAD on a sphere retains the time complexity $O(n \log(n))$ of the planar variant. In our case, this enabled computing scores of $\approx40\,000$ pairs of $\approx 9$~km global fields, each consisting of more than 6 million points, in less than three days by utilizing 25 cores on a single computer. The native IFS model grid was used for comparisons, thus avoiding interpolations that can deteriorate the fields and smooth spikes. 

When interpreting the PAD value as a location error, we must bear in mind that other types of forecast errors beyond the displacement of features can occur, for instance, due to systematic model biases or discrepancies in intensity, spatial extent, or duration of the precipitation. In these cases, the algorithm does not have a way to separate truly displaced precipitation features from features that do not have a corresponding counterpart (a human would also probably struggle with this somewhat subjective task), and some features could end up being attributed to unrelated precipitation features far away. Implementing a distance cut-off threshold partly mitigates this problem and ensures that regional biases (e.g., a particular wet forecast over a certain region) do not end up being accounted as large displacement errors. We selected a cut-off value of 3000 km, ensuring that regional errors more than 3000 km apart do not get attributed as large displacements. It might still be possible to develop smarter ways to filter the attributions for unrealistic assignments (e.g., instead of a hard cutoff rule that simply disregards all attributions above a certain threshold distance, the influence of the unrealistically distant attributions could be managed in a less rigid manner by applying a distance-dependent weight when calculating the PAD value), but this has not been investigated further here. 

Due to this cut-off rule and also because forecasts can have global biases, there will always be some unattributed precipitation remaining that informs on the overall and regional errors in terms of magnitude. We strongly recommend looking at the non-attributed precipitation together with the PAD distance results when evaluating forecasts with this method. For example, when comparing two forecasts, and both parameters, the PAD value and the amount of non-attributed precipitation, are better for the first forecast (i.e., have lower values), then one can reasonably conclude that the first forecast is better and vice versa. However, if one parameter is better for the first forecast and the other for the second forecast, it is not possible to conclude which forecast is better. There is also a third option, that the value of one parameter is roughly the same for both forecasts, while the other one is not the same. In this case, one can also conclude that the forecast with the better value of the second parameter is better.

In the results section, we have showcased how the PAD-on-sphere method can be employed to analyze single cases, evaluate forecast quality over a long period, assess forecast improvements over time, and spot problematic regions or locations. In particular, we have tested the quality of the operational ECMWF forecasts in the period 2016--2023. Due to the lack of availability of high-resolution, high-quality global precipitation observations, we employed the short-term forecasts by the same model as verification truth. Interpretation of the verification results should be careful and take into account the strong dependence between IFS and the pseudo-observations. 

The summary results for the whole period (Fig.\ref{fig:IFS_bar_graph}) show how location errors in the IFS model grow steadily with increasing lead time for all regions. At the same time, the unattributed precipitation grows, indicating global and regional biases and intensity errors that are not considered in the location error. An analysis with a conventional metric such as the RMSE (Fig.\ref{fig:IFS_bar_graph}d), contrarily, shows that RMSE already has large values in the first lead times, and they do not grow much further. This is indicative of small location errors that are already penalized on the first lead times by the RMSE and are not penalized further when the displacements grow larger. Another striking difference between the PAD results and the traditional verification measure is that the scores are much more similar across regions for PAD than for RMSE. RMSE is very much affected by the absolute amounts of precipitation in each region, whereas for the PAD, the absolute precipitation quantities are just used to weigh the contributions of the different displacements, resulting in more comparable numbers across regions. The regional analysis shows that for short lead times, the location errors in the Tropics are larger than those in the Midlatitudes, but conversely, at longer lead times, the situation is reversed with larger errors in the Midlatitudes. This is also clearly seen in the local diagnostics (Fig.\ref{fig:IFS_LPAD_global}) or even for a single valid date (Fig.\ref{fig:IFS_PAD_example}) and is in line with the fact that extratropical cyclones can move long distances over 9 days whereas the position of the ITCZ and associated convective precipitation does not change so much. While the precipitation could still move longitudinally across the ITCZ, slowly evolving processes such as the Madden-Julian Oscillation, the Monsoons, or the positioning of the Walker cell constrain the areas where rainy cells can move over a period of 9 days.

The global scores show a steady evolution or a slight reduction of location errors over the seven-year period (e.g. on the order of 1 km per year for the 3- and 5-days-ahead forecasts, Fig.\ref{fig:IFS_timeseries_seasonal}). The time evolution over extratropical regions also displays a strong annual cycle, with larger errors during the seasons with more prevalence of convective precipitation, as one would expect. A close look over the LPAD values over Europe or the Maritime Continent reveals that orographically enhanced precipitation results in lower PAD values over topographic features. This indicates that IFS forecasts do a good job over orography and that the method is indeed measuring the location error because when orography forces the precipitation, the location is determined by the orography itself and not by the more uncertain position of a convective cell or a passing front.

In summary, we have shown that it is possible to estimate location errors in global precipitation forecast by attributing as much precipitation as possible from any grid point in the forecast field to a corresponding close-by rainy point in the observations and then measuring the average length of all the attributions. Beyond the usages we have shown above, this technique will be useful to compare the quality of precipitation forecasts from different models, model versions, or parameter configurations, something which is often overlooked in favor of other dynamical model parameters that are easier to verify and less sensitive to location errors. Since the time complexity of the algorithm is constrained, we believe computing location errors for higher-resolution global forecasts, such as the 4.4 km and 2.8 km simulations being developed under the Destination Earth (DestinE) program, should still be feasible. Classical non-spatial metrics such as RMSE, contrarily, would be even more contaminated by double penalty effects at these resolutions.

Optimal transport opens possibilities for forecast verification of other quantities as well, although this work has been motivated specifically by precipitation and might require special care when applying it to other settings. Beyond deterministic predictions, most forecast-producing centers also run ensembles to derive probabilistic predictions. While ensemble-based forecasting does help to alleviate the double penalty issue, it does not completely resolve it, and thus adapting the methodology for probabilistic predictions would potentially also be of interest.

\section*{Acknowledgements}
We are grateful to Thomas Haiden and Richard Forbes for fruitful discussion on the methodology, results and their interpretation. We want to thank Willem Deconinck for assisting with the computation of the IFS grid box area. 

\section*{Funding}
Slovenian Research And Innovation Agency (Javna agencija za znanstvenoraziskovalno in inovacijsko dejavnost RS) research core funding No. P1-0188. The work presented in this paper has been produced in the context of the European Union’s Destination Earth Initiative and relates to tasks entrusted by the European Union to the European Centre for Medium-Range Weather Forecasts implementing part of this Initiative with funding by the European Union. Views and opinions expressed are those of the author(s) only and do not necessarily reflect those of the European Union or the European Commission. Neither the European Union nor the European Commission can be held responsible for them.

\section*{Data Availability Statement}
A Python software package for efficient calculation of the PAD on the sphere is freely available on GitHub (\url{https://github.com/skokg/PAD_on_Sphere}) and Zenodo (\url{https://doi.org/10.5281/zenodo.15180114}).

\section*{Author Contributions}
\textbf{Gregor Skok}: conceptualization, data curation, formal analysis, funding acquisition, investigation, methodology, resources, software, validation, visualization, writing – original draft, writing – review \& editing. \textbf{Llorenç Lledó}: data curation, investigation, methodology, validation, writing – original draft, writing – review \& editing.

\bibliography{references}

\end{document}


\maketitle

\makeatletter 

\renewcommand{\thefigure}{S\@arabic\c@figure}
\renewcommand{\thesection}{S\@arabic\c@section}
\renewcommand{\thetable}{S\@arabic\c@table}
\makeatother

\section{IFS grid box area and precipitation}

\begin{figure}[h]
    \centering
    \includegraphics[width=5cm]{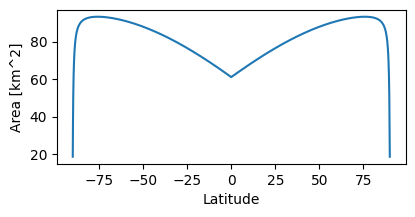}
    \caption{Area of the IFS O1280 grid boxes as a function of latitude.}
    \label{fig:gridarea}
\end{figure}

\begin{figure}[h]
\centering
\includegraphics[width=10cm]{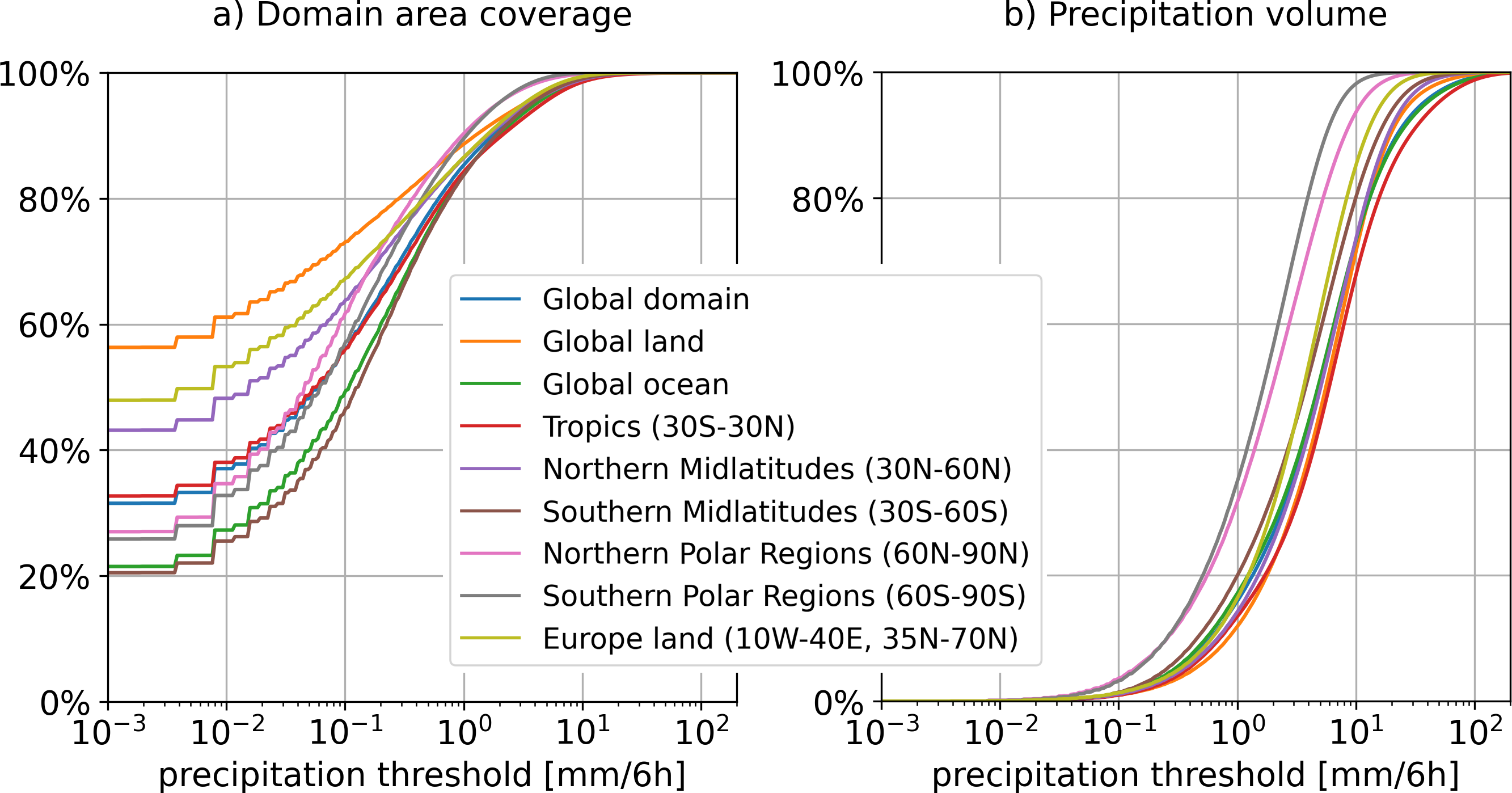}
\caption{a) The average portion of the domain area covered by precipitation below a certain threshold. b) The average portion of the precipitation volume in a domain contributed by precipitation below a certain threshold. In both subfigures, the values are averaged over 7 years of the 6-hourly precipitation fields of pseudo-observations.} 
\label{fig:regional_statistics}
\end{figure}

\clearpage
\section{Bias and Non-Attributed Precipitation PDF}

\begin{figure}[h]
\centering
\includegraphics[width=11cm]{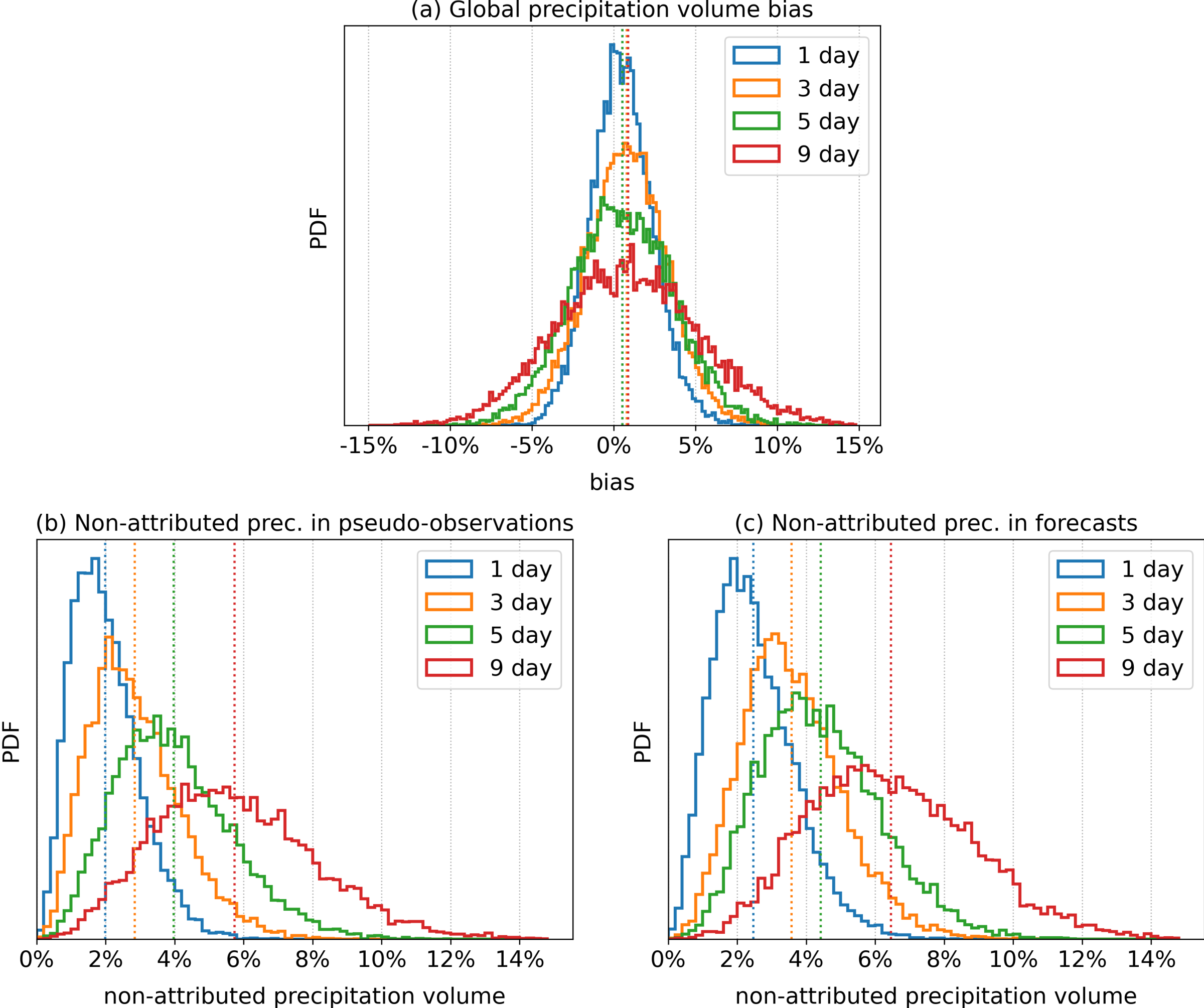}
\caption{(a) The Probability Density Function of the global precipitation volume bias (the difference between the sums of precipitation volume in the forecasts and pseudo-observations, divided by the sum of precipitation volume in the pseudo-observations, expressed as a percentage) in the 7-year period for the 1-, 3-, 5-, and 9-day-ahead forecasts. (b,c) The PDF of the Non-Attributed precipitation fraction (the sum of non-attributed precipitation volume divided by the sum of all precipitation in the field, expressed as a percentage) in the pseudo-observations and the forecasts for the 7-year period for the 1-, 3-, 5-, and 9-day-ahead forecasts using a 3000 km distance cutoff.}
\label{fig:bias_and_nap_PDF_}
\end{figure}

\clearpage
\section{Cutoff implementation comparison}

The cutoff can be implemented during the attribution phase or via post-processing. In the first case, when the nearest neighboring point in the other field is identified at each attribution step, a check is made if the distance is larger than the cutoff distance. If this is true, the point is removed from the list of remaining points, and its remaining precipitation will remain unattributed till the end. The point in the other field is not removed and can potentially still be attributed at some later step.

Alternatively, the cutoff could be implemented after the attribution phase finishes (without using any cutoff distance) as a post-processing step, with all the attributions with a distance larger than the cutoff disregarded. In this case, there will be more non-attributed precipitation compared to the first approach since both fields are affected when a certain attribution is disregarded. 

\begin{figure}[btp]
\centering
\includegraphics[width=0.95\textwidth]{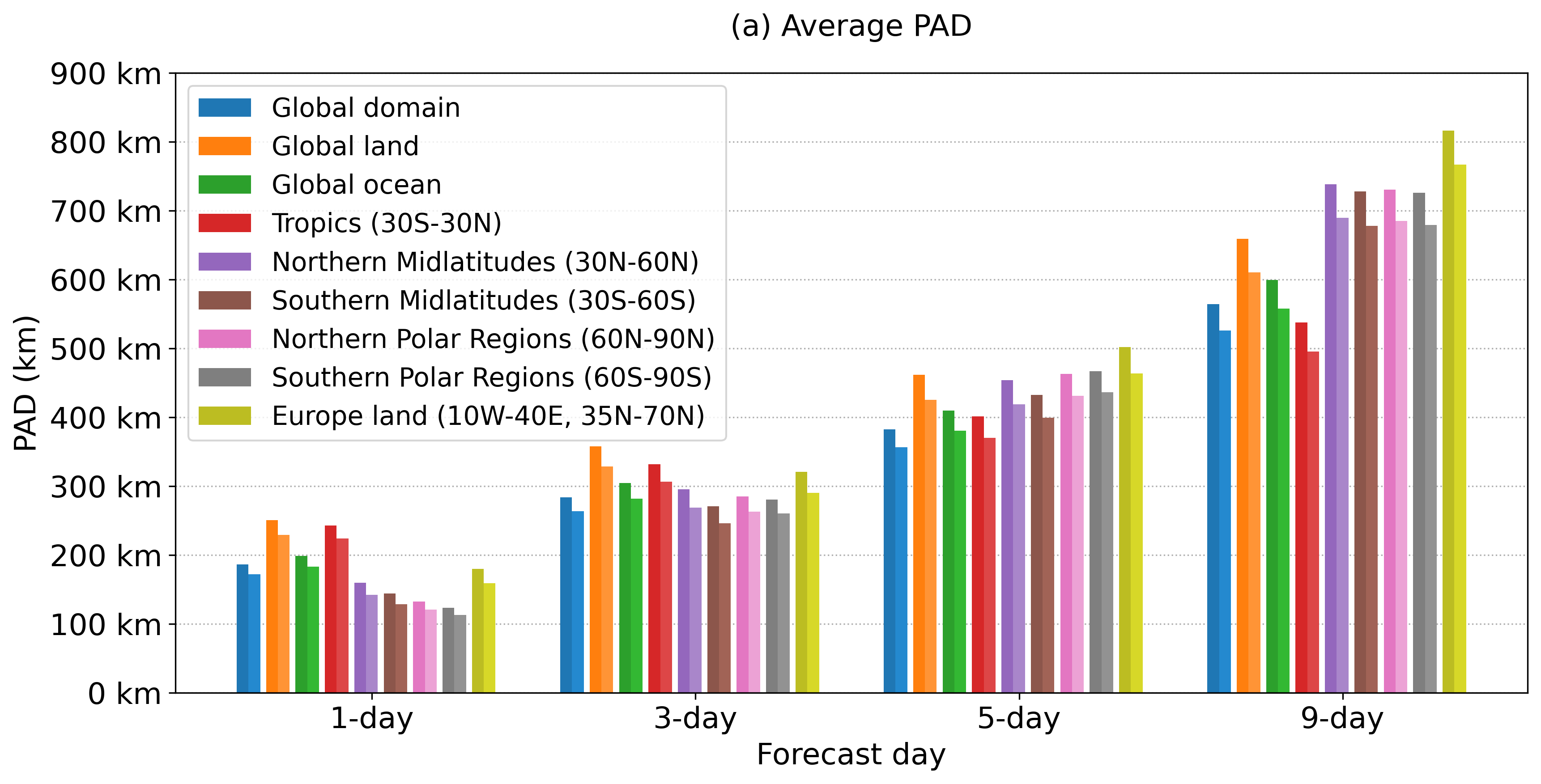}
\includegraphics[width=0.95\textwidth]{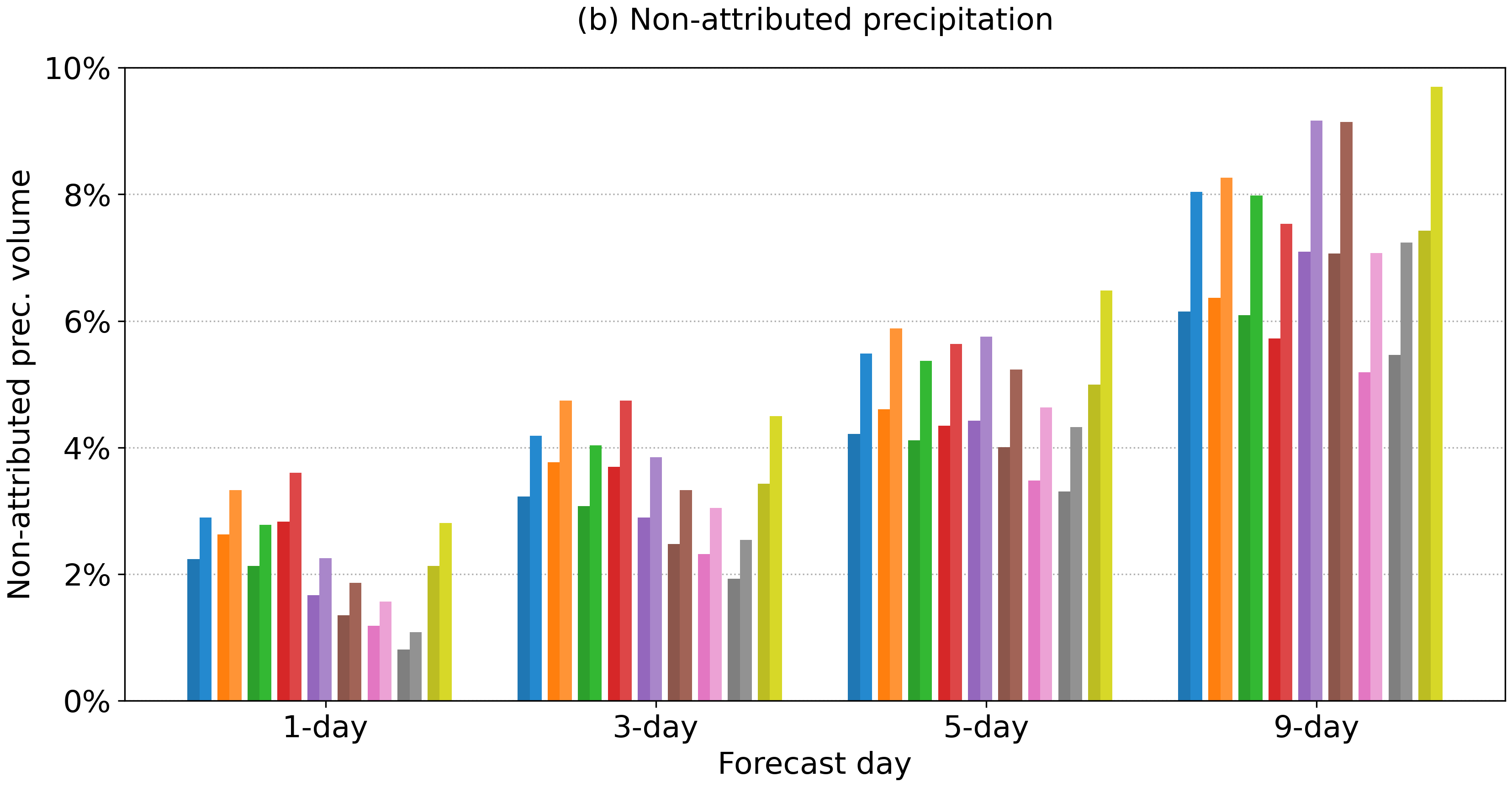}
\caption{(a) The average PAD value in the 7-year period for different sub-regions and two implementations of a 3000 km cutoff distance. (b) The average amount of non-attributed precipitation (the sum of non-attributed precipitation volume in both fields divided by the sum of all precipitation in both fields, expressed as a percentage) for different sub-regions and two implementations of a 3000 km cutoff distance. In both sub-figures, each group of two similarly colored bars represents the results for the cutoff implementation during the attribution phase (on the left side) and  the implementation via post-processing (on the right side)}  
\label{fig:IFS_cutoff_implementation}
\end{figure}

Fig.\ref{fig:IFS_cutoff_implementation} shows a comparison of the two approaches for the Mean PAD and Non-attributed precipitation of the IFS forecasts in the 7-year period. Regarding the PAD values, the approach with the post-processing produces a bit smaller mean PAD values for all forecast times and sub-regions (the difference is about 8\%). On the other hand, as expected, the post-processing-based approach consistently produces more non-attributed precipitation for all forecast times and sub-regions. The relative difference in the amount of non-attributed precipitation is about 30\%.

\clearpage
\section{Cutoff distance sensitivity analysis}

The cutoff distance sensitivity analysis was performed by comparing the results of the 3000 km cutoff with the results using a pair of smaller and larger cutoff values (i.e., 2000 and 4000 km)

\begin{figure}[btp]
\centering
\includegraphics[width=0.95\textwidth]{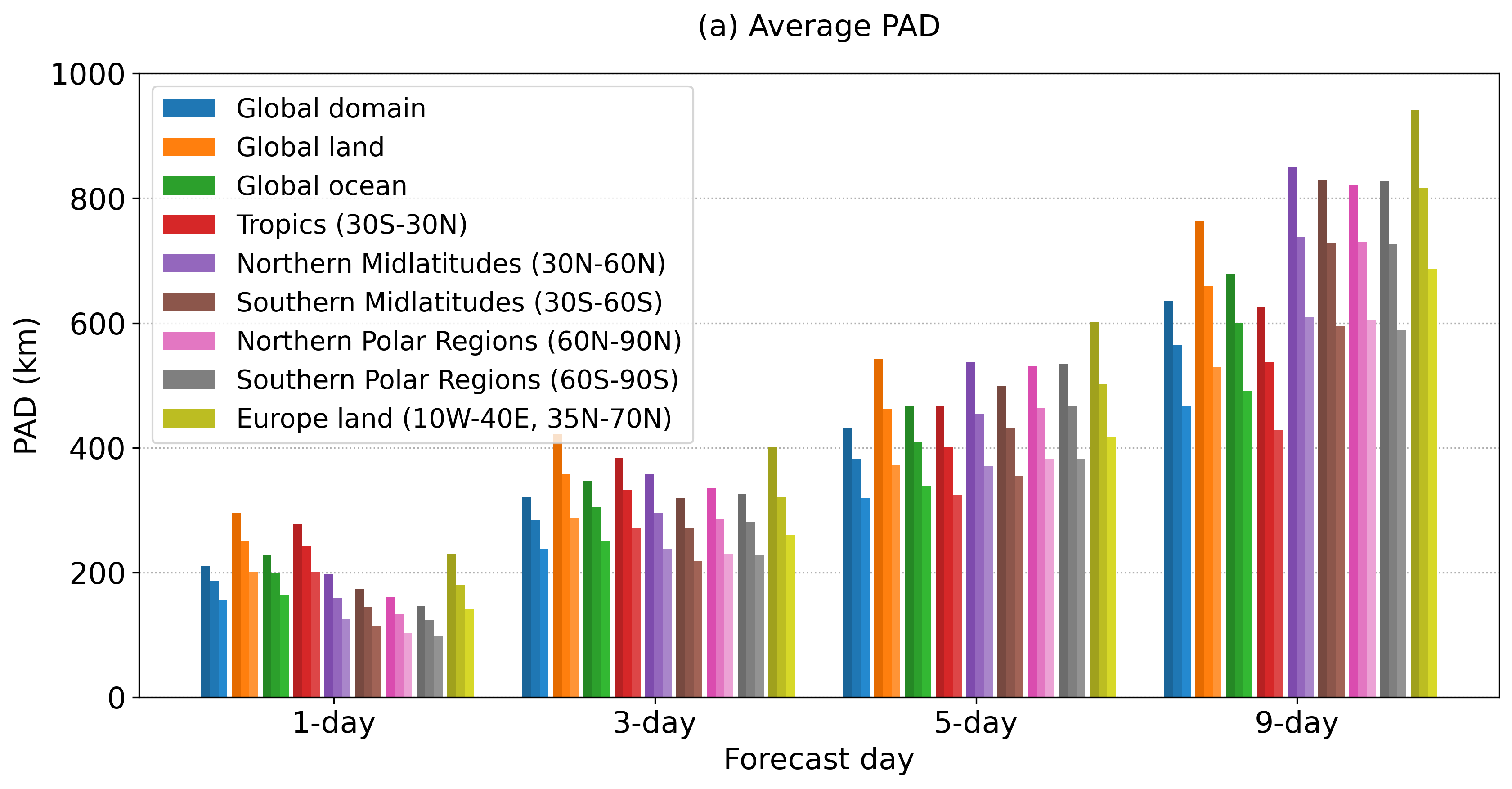}
\includegraphics[width=0.95\textwidth]{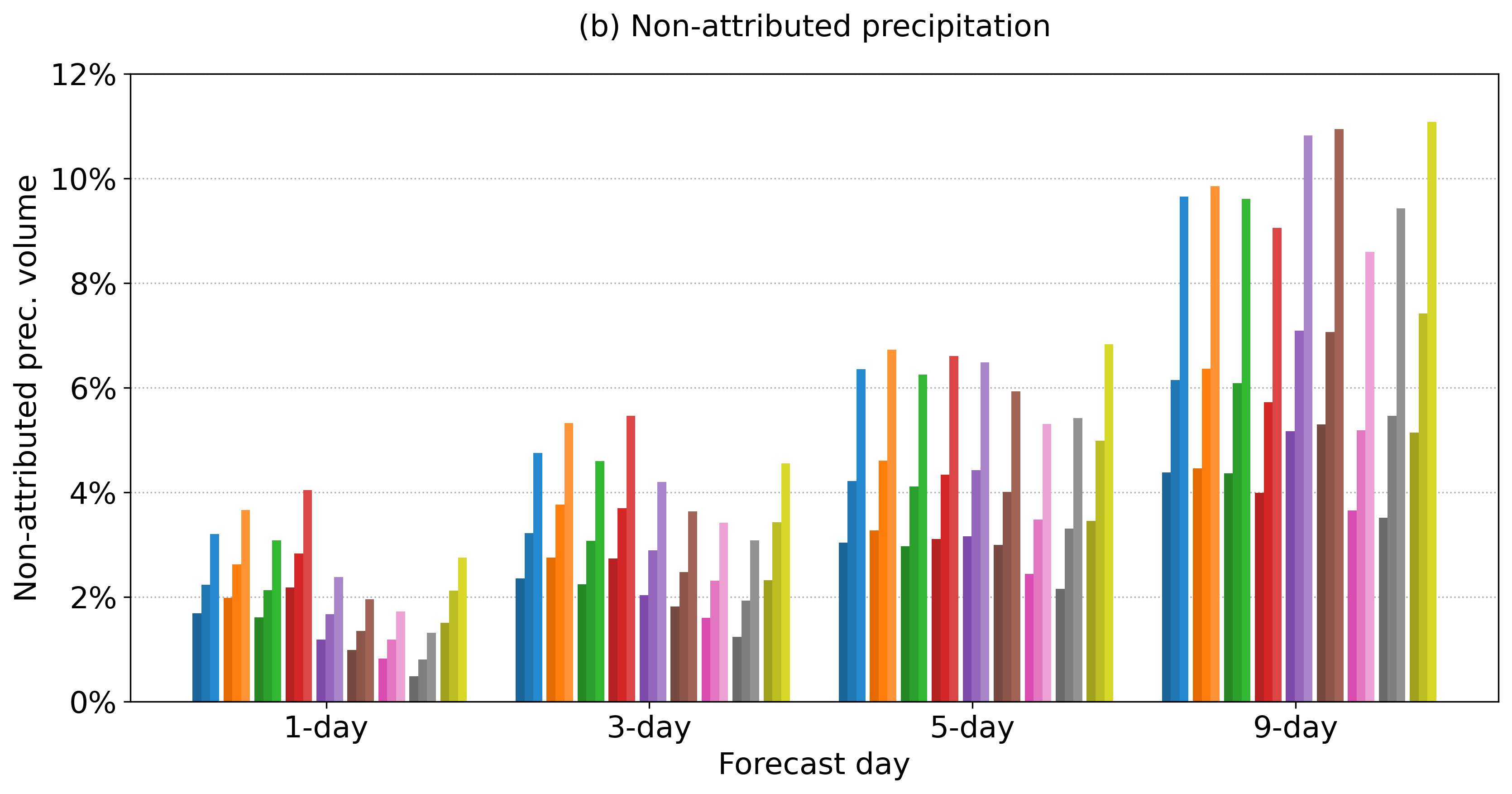}
\caption{(a) The average PAD value in the 7-year period for different sub-regions and three values of cutoff distance. (b) The average amount of non-attributed precipitation (the sum of non-attributed precipitation volume in both fields divided by the sum of all precipitation in both fields, expressed as a percentage) for different sub-regions and three values of cutoff distance. In both subfigures, each group of three similarly colored bars represents the results for the three cutoff distances of 4000, 3000, and 2000 km, from left to right, respectively.}  
\label{fig:IFS_cutoff_sensitivity}
\end{figure}

Fig.\ref{fig:IFS_cutoff_sensitivity}a shows the average PAD value in the 7-year period for different sub-regions. As expected, a smaller cutoff distance always results in a smaller mean PAD value, with the results for the 4000 and 2000 km cutoffs being about 15\% larger or smaller compared to the 3000 km, respectively. At the same time, although the PAD values change if a different cutoff is used, the ranking of results with respect to different forecast times and sub-domains does not change (e.g., for the 1-day-ahead forecasts, the Tropics have the largest PAD value of all the latitudinally delineated sub-regions, while for the 9-day-ahead forecast, they have the smallest value). 

Fig.\ref{fig:IFS_cutoff_sensitivity}b shows the average fraction of non-attributed precipitation. As expected, the amount of non-attributed precipitation increases if a smaller cutoff distance is used, as more precipitation remains unattributed. On average, for the 4000 km cutoff, there is about 30\% less non-attributed precipitation compared to the 3000 km cutoff. Contrary, for the 2000 km cutoff, there is about 50\% more non-attributed precipitation compared to the 3000 km cutoff. For the 9-day-ahead forecasts, the fraction of the non-attributed precipitation is around 10\% for the 2000 km cutoff, meaning that a significant portion of the precipitation is not attributed. On the other hand, for the 3000 km cutoff, this fraction is only around 6\%. The considerable influence of the cutoff on the non-attributed precipitation is not surprising, since the area of the spherical cap to which the precipitation at a certain location can be attributed varies with the squared value of the cutoff distance. For example, in the case of the 2000 km cutoff, the area size of this region is less than half of the region for the 3000 km cutoff and only about a quarter of the 4000 km cutoff region. 

\begin{figure}[btp]
\centering
\includegraphics[width=0.95\textwidth]{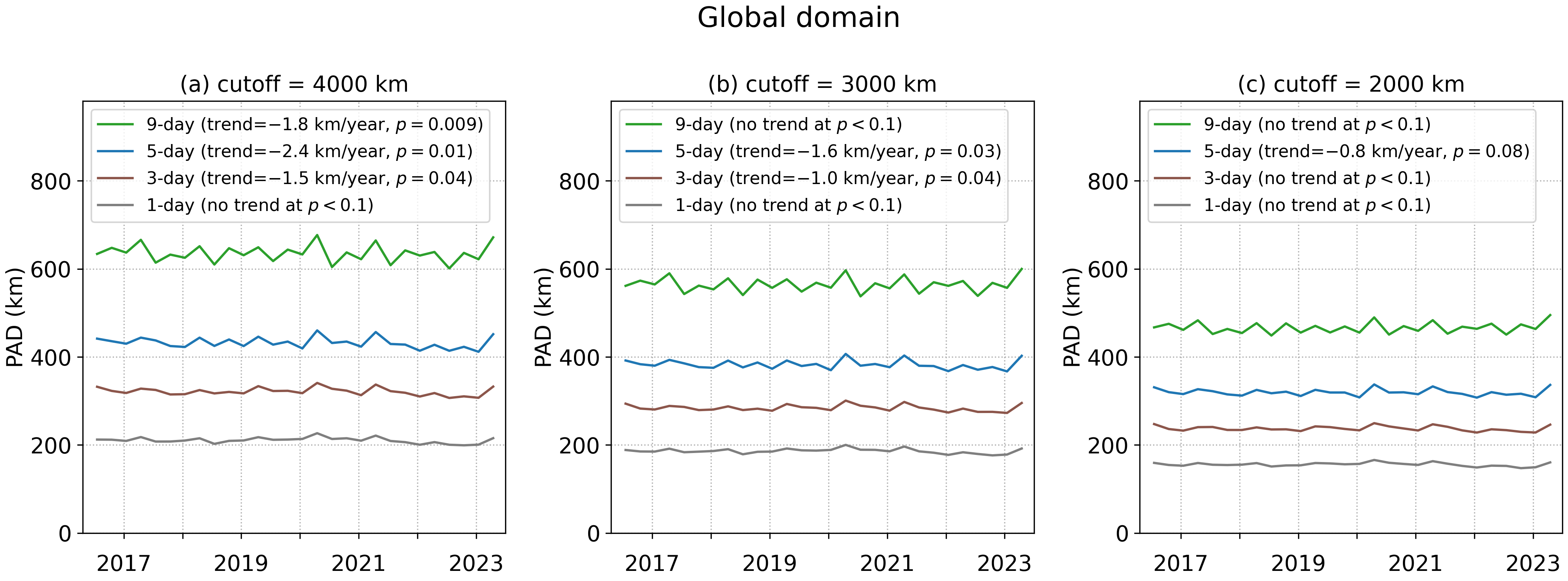}

\vspace{0.5cm}

\includegraphics[width=0.95\textwidth]{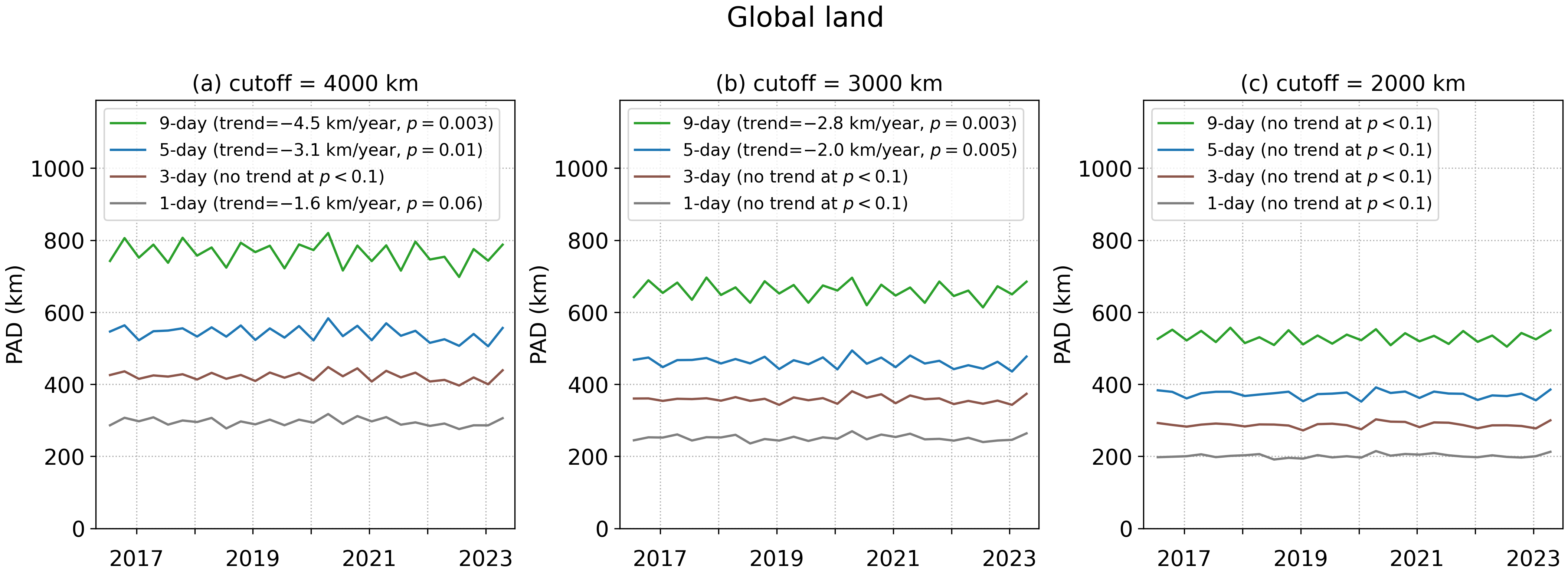}

\vspace{0.5cm}

\includegraphics[width=0.95\textwidth]{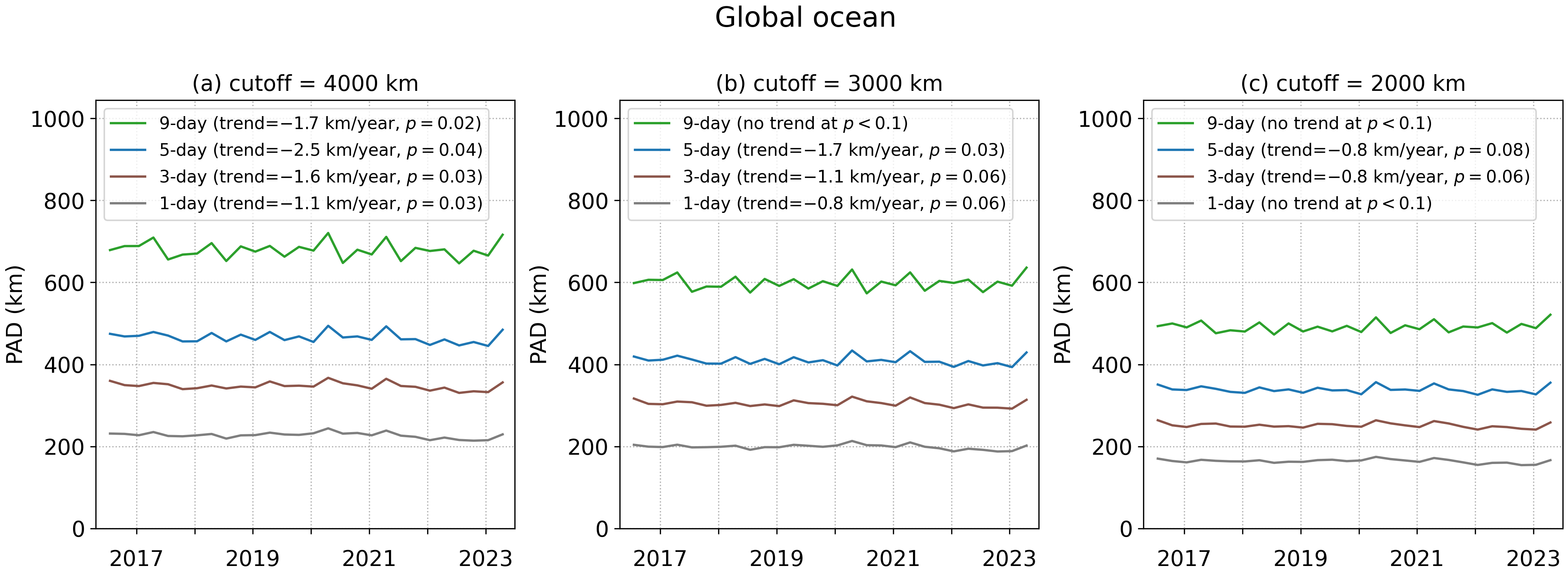}
\caption{Timeseries of mean seasonal PAD values for the (top) Global, (middle) global land and (bottom) global ocean sub-regions, using three values of cutoff distances: (a) 4000 km, (b) 3000 km, and (c) 2000 km. Statistically significant trends are indicated in the legend. A significance level of $p < 0.1$ has been used for all tests.}  
\label{fig:IFS_cutoff_sensitivity_00}
\end{figure}

\begin{figure}[btp]
\centering
\includegraphics[width=0.95\textwidth]{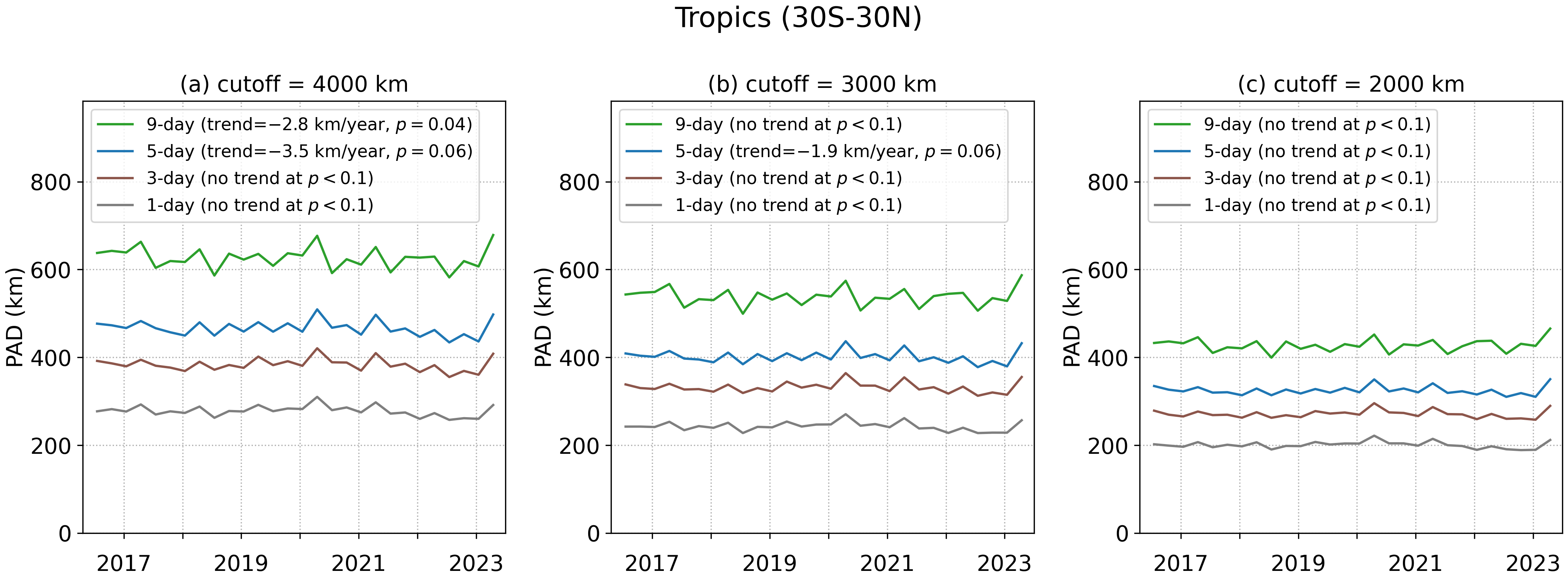}

\vspace{0.5cm}

\includegraphics[width=0.95\textwidth]{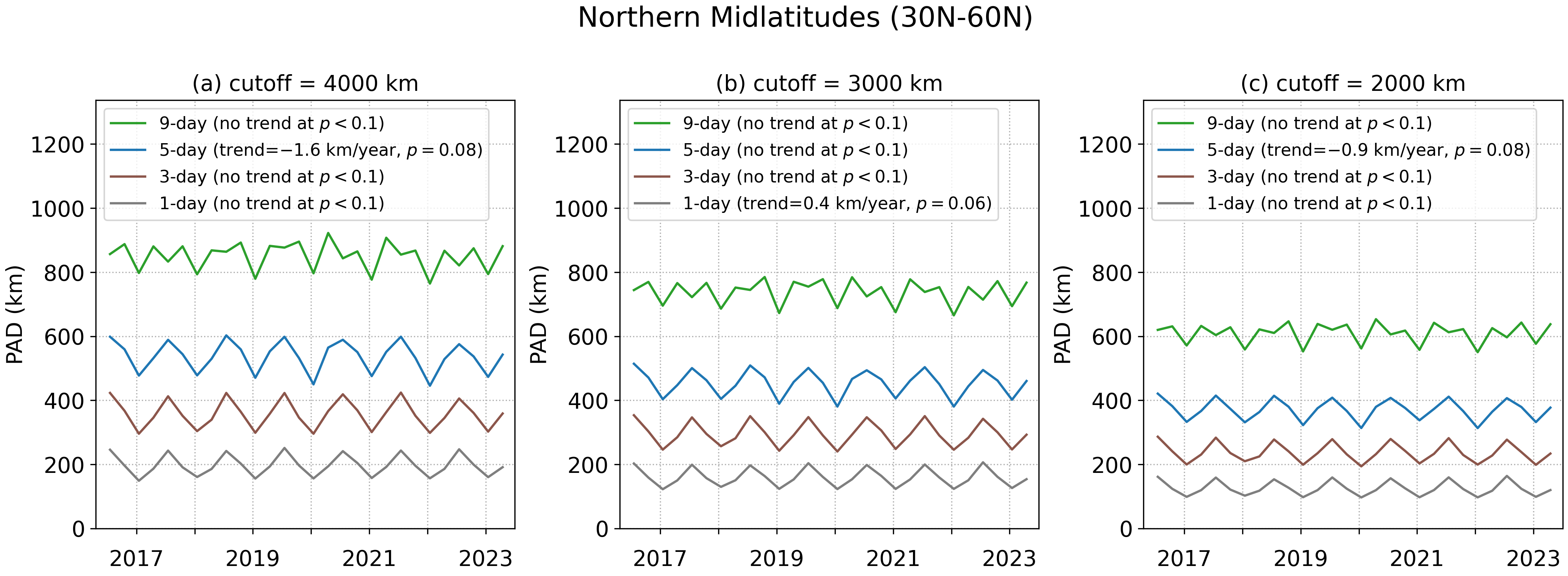}

\vspace{0.5cm}

\includegraphics[width=0.95\textwidth]{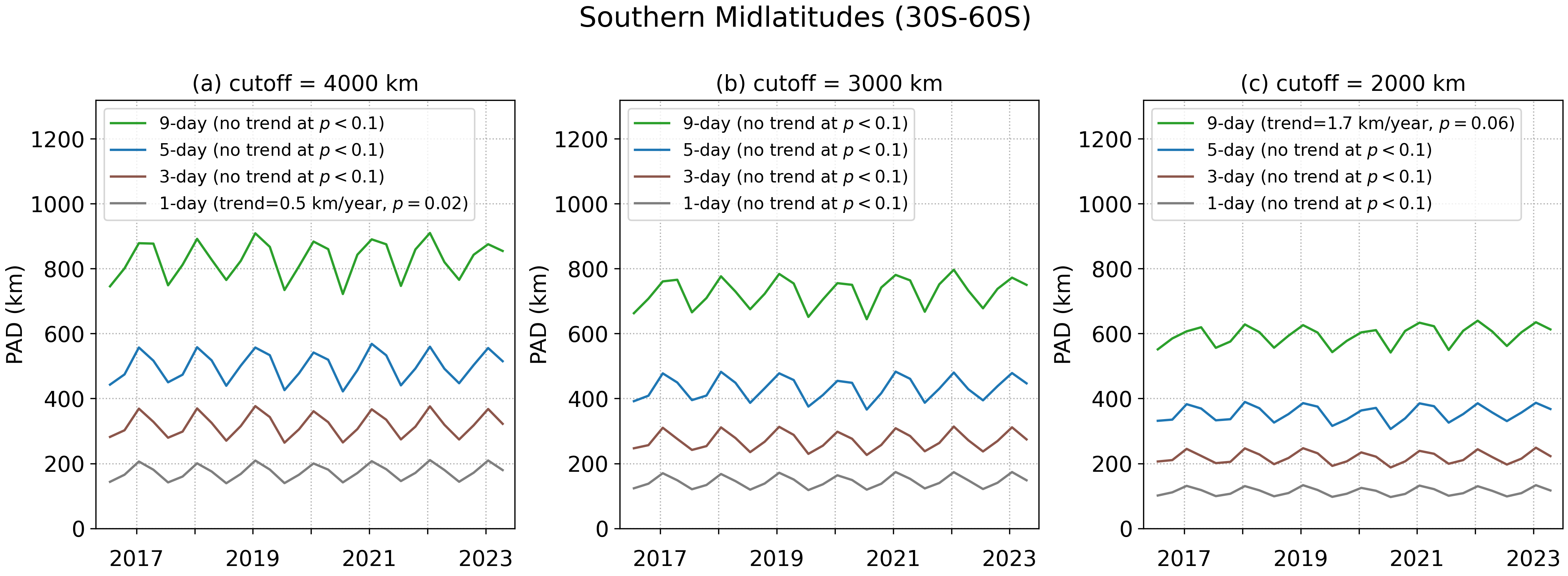}
\caption{Same as Fig.\ref{fig:IFS_cutoff_sensitivity_00} but for the Tropics and Northern and southern Midlatitudes.}  \label{fig:IFS_cutoff_sensitivity_03}
\end{figure}

\begin{figure}[btp]
\centering
\includegraphics[width=0.95\textwidth]{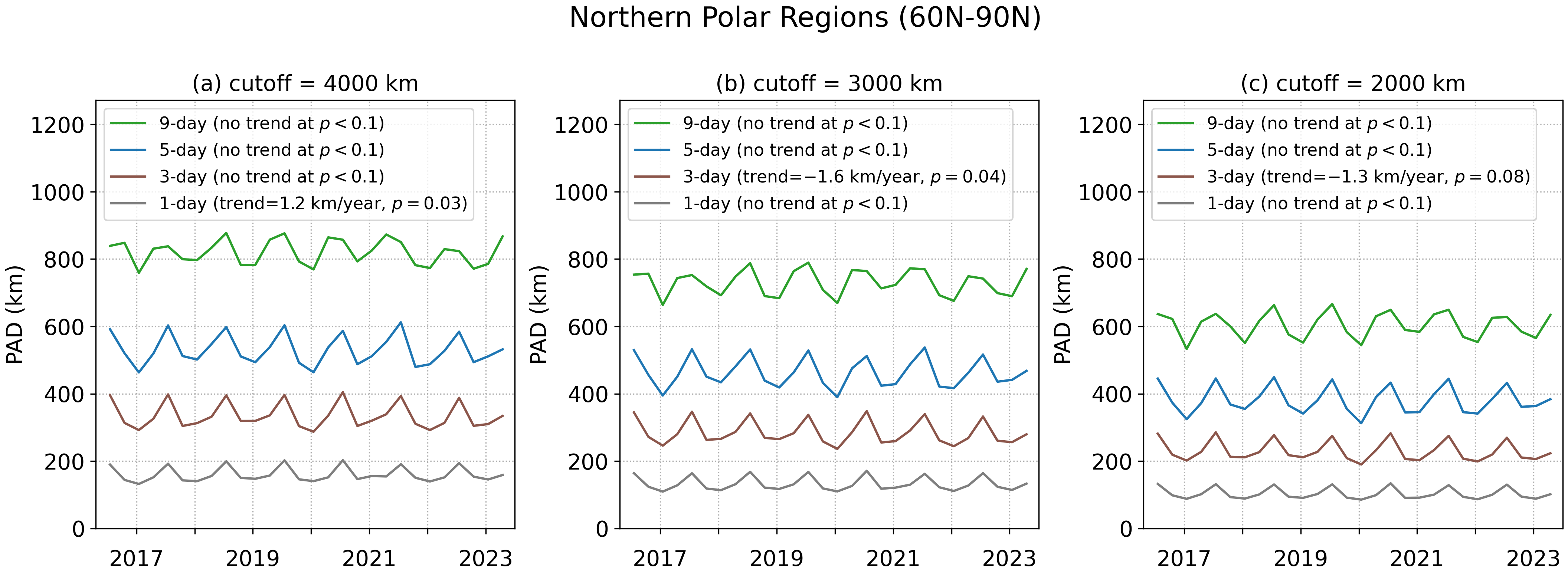}

\vspace{0.5cm}

\includegraphics[width=0.95\textwidth]{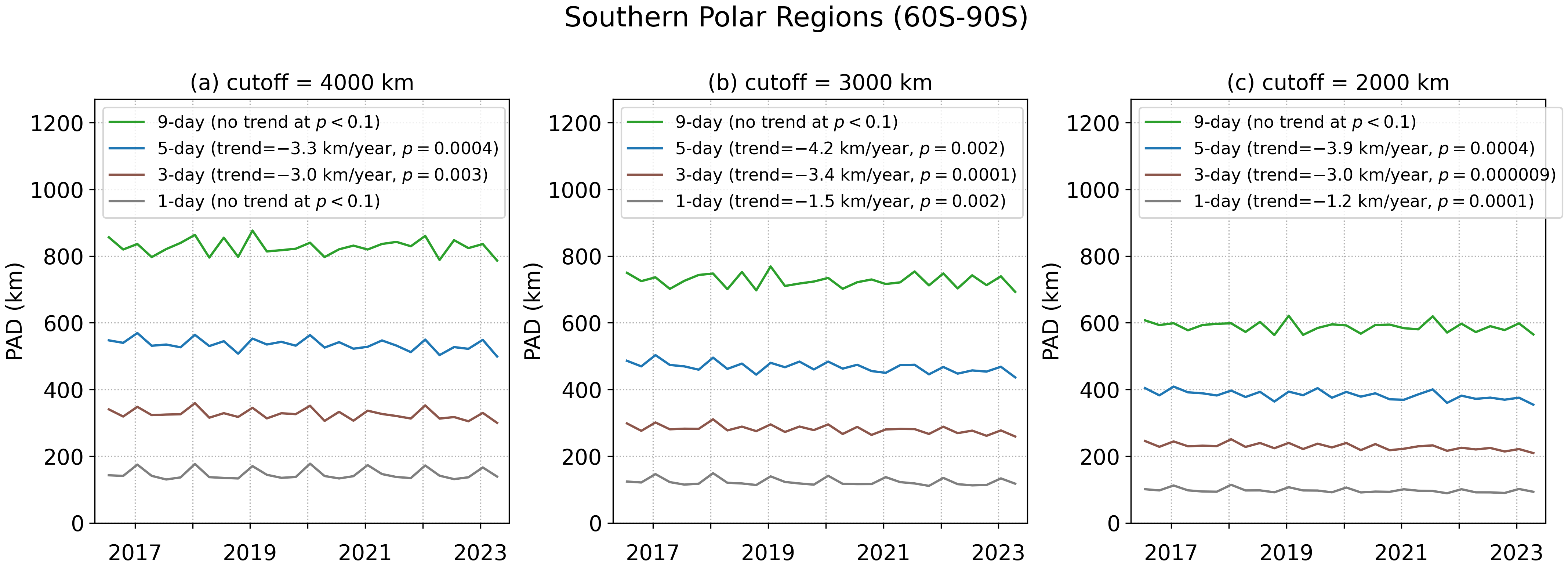}

\vspace{0.5cm}

\includegraphics[width=0.95\textwidth]{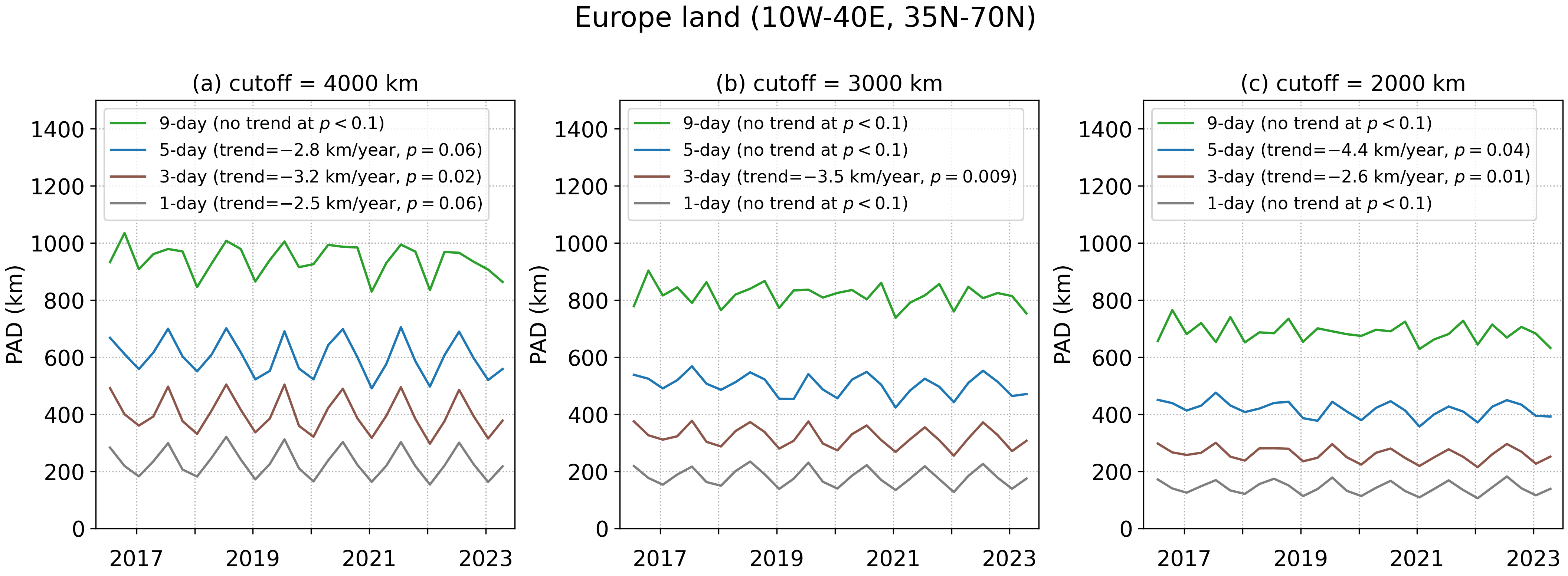}
\caption{Same as Fig.\ref{fig:IFS_cutoff_sensitivity_00} but for the  Northern and Southern Polar Regions and Europe land sub-region.}
\label{fig:IFS_cutoff_sensitivity_06}
\end{figure}

Figs.\ref{fig:IFS_cutoff_sensitivity_00}-\ref{fig:IFS_cutoff_sensitivity_06} show the time evolution of mean seasonal PAD values for different sub-regions for the three cutoff distances. In some sub-regions, a cutoff does not have a large influence on the statistical significance and sign of the trends (e.g., the Northern \& Southern Midlatitudes and the Northern Polar Regions that, in most cases, do not exhibit statistically significant trends). In other cases, the influence can be substantial. For example, for the Europe-land sub-region, where there can be between one and three statistically significant decreasing trends depending on the cutoff being used.  Notably, there are no examples in any sub-region where a trend changes its sign when a different cutoff distance is used (i.e., that a statistically significant negative trend would become a positive trend or vice versa).

\clearpage
\section{Hedging}\label{sec:bias_hedging}

There are two simple hedging strategies that can be used to increase or decrease the amount of precipitation in a forecasted field: multiplying all the values by a constant (hedging by multiplication) or increasing/decreasing all the values by a constant (hedging by addition/subtraction). Hedging by multiplication will only change the values where the precipitation is larger than zero, with the larger values increasing or decreasing more than the smaller values. Hedging by addition/subtraction will increase/decrease the values everywhere for the same amount (provided the resulting value is larger than zero - otherwise, it is set to zero). In the case of addition, there will not be any points with zero precipitation left. In the case of subtraction, the number of zero-valued points might increase.

\begin{figure}[bp]
\centering
\includegraphics[width=0.9\textwidth]{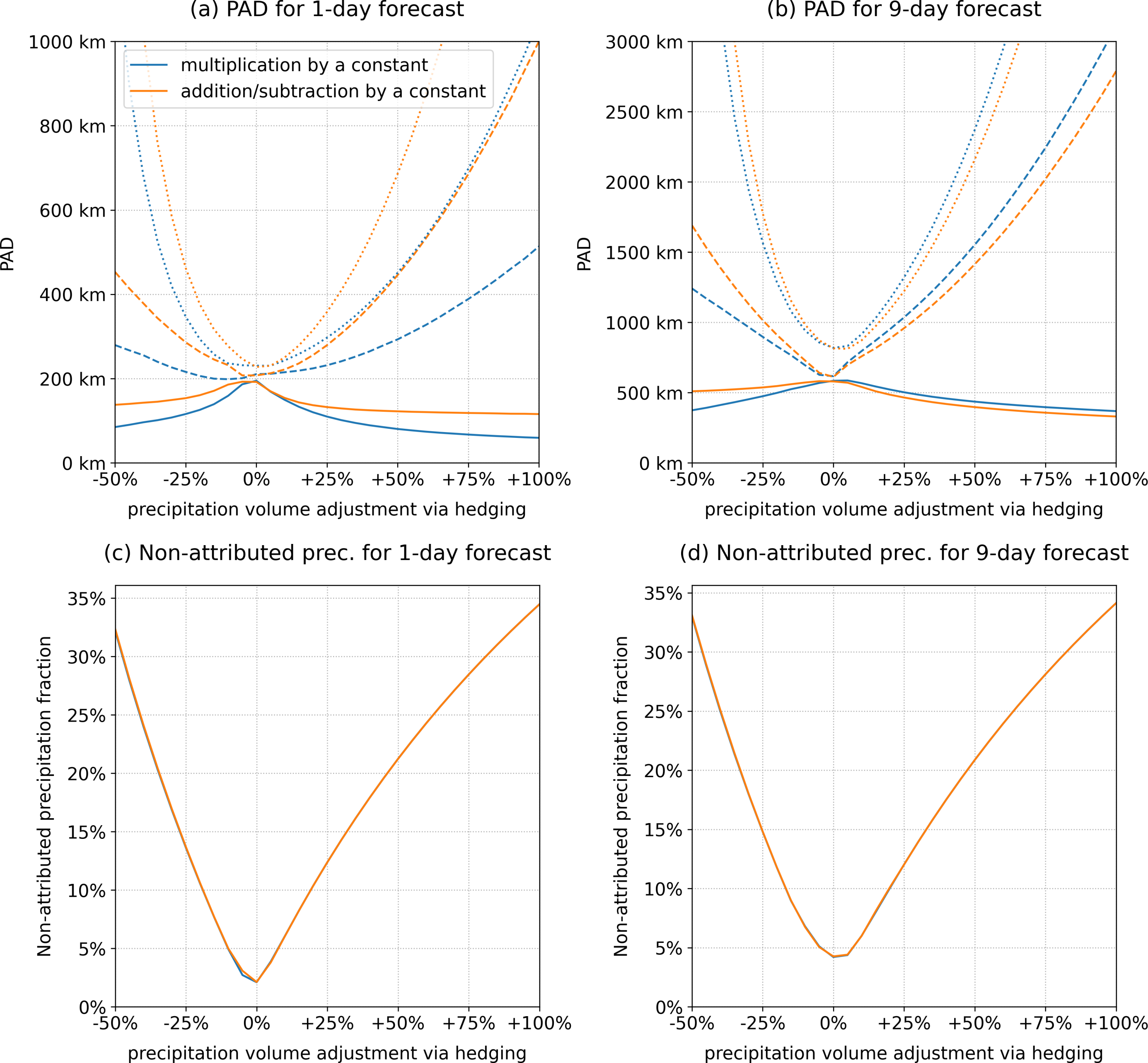}
\caption{The influence of hedging on the PAD results. (a,b) The influence of precipitation volume adjustment via hedging on the PAD value for the 1-day (left) and 9-day (right) IFS forecast of 00-06 UTC precipitation for 11 October 2022. The solid lines represent the original PAD values, the dashed lines represent the corrected PAD values calculated using Eq.\ref{eq:PADcorr} with $p=3$, and the dotted lines the corrected PAD values using Eq.\ref{eq:PADcorr2} with $p=4$. (c,d) The Non-attributed precipitation fraction (the sum of the non-attributed precipitation volume in both fields divided by the total precipitation volume in both fields). The non-attributed precipitation fraction for both hedging strategies is almost the same, with the blue curve being beneath the orange curve.}
\label{fig:out_IFS_bias_hedging}
\end{figure}

Fig.\ref{fig:out_IFS_bias_hedging}(a-b) shows the influence of hedging on the PAD value for the 1- and 9-day-ahead IFS forecast of 00-06 UTC precipitation for 11 October 2022 (the same case is shown in Fig.5 in the main manuscript). As can be observed, both hedging strategies result in smaller PAD values, indicating a better forecast, but, at the same time, increase the amount of non-attributed precipitation, indicating a worse forecast (Fig.\ref{fig:out_IFS_bias_hedging}(c-d)). Interestingly, in the case of positive hedging (increasing the total amount of precipitation), a lower PAD value for the 1-day-ahead forecast is produced using hedging by multiplication, while hedging by addition produces a slightly smaller PAD value for the 9-day-ahead forecast. This behavior probably occurs due to the precipitation in the 1-day-ahead forecast mostly being at the correct locations, so multiplying by a constant increases the amount of precipitation in the forecast at mostly correct locations, while hedging by addition also increases the precipitation at other locations. For the 9-day-ahead forecasts, there is less overlap of precipitation in the two fields, and multiplying by a constant thus does not perform better than hedging by addition. In the case of negative hedging (decreasing the total amount of precipitation),  the hedging by multiplication produces a lower PAD value for both the 1- and 9-day-ahead forecasts.

If the possibility of hedging is a concern, there are some options on how to avoid or mitigate it. One option is to check the forecast bias or the amount of non-attributed precipitation in addition to the PAD value, and if the bias (or the amount of non-attributed precipitation) is large, declare the forecast not good even though the PAD value might actually be good. A second option is to normalize the fields before calculating the PAD - in this case, hedging via multiplication is not possible, but global normalization can induce regional biases into the forecast. 

Another option is to define a corrected PAD value, which would also take into account the bias or the non-attributed precipitation. For example, a bias-corrected PAD value can be defined as 
\begin{equation}
PAD_\textrm{corr} =  PAD \cdot (1+bias)^p,
\label{eq:PADcorr}
\end{equation}
where $PAD$ is the original PAD value, $bias=\frac{|o - f|}{o}$ with $o$ and $f$ being the sums of precipitation volume in the pseudo-observations and the forecast, respectively, and $p$ is the power parameter. Since $bias \geq 0$, the $PAD_\textrm{corr}$ can only be equal or larger than $PAD$.

Alternatively, a correction can also be based on the amount of non-attributed precipitation. For example, 
\begin{equation}
PAD_\textrm{corr} =  PAD \cdot (1+napf)^p,
\label{eq:PADcorr2}
\end{equation}
where $napf= \frac{o_{nap}+f_{nap}}{attr}$ with $o_{nap}$ and $f_{nap}$ being the sums non-attributed precipitation precipitation volumes in the pseudo-observations and the forecast, respectively, and $attr$ is the sum of attributed precipitation volume (which will be the same for both fields). 

Fig.\ref{fig:out_IFS_bias_hedging} shows the corrected PAD values calculated using Eqs.\ref{eq:PADcorr} and \ref{eq:PADcorr2}. The corrected values increase as the amount of precipitation artificially increases or decreases when applying any of the two hedging strategies, meaning hedging does not produce better results. However, it is worth noting that these results are based on an analysis of a single case, and a more thorough analysis, preferably one based on many cases, would be needed to more reliably determine the influence of hedging on PAD.

\clearpage
\section{Influence of forecast smoothness}

One interesting aspect of a metric's behavior is how its value changes if the forecasted field is smoothed. Two situations were analyzed to gain some basic insight into this behavior: a simple idealized case and a real-world one. In both cases, one of the fields was not smoothed, while the other was smoothed using a circular smoothing kernel with a prescribed diameter, conserving the total amount of precipitation in the field.

\begin{figure}[bp]
\centering
\includegraphics[width=0.95\textwidth]{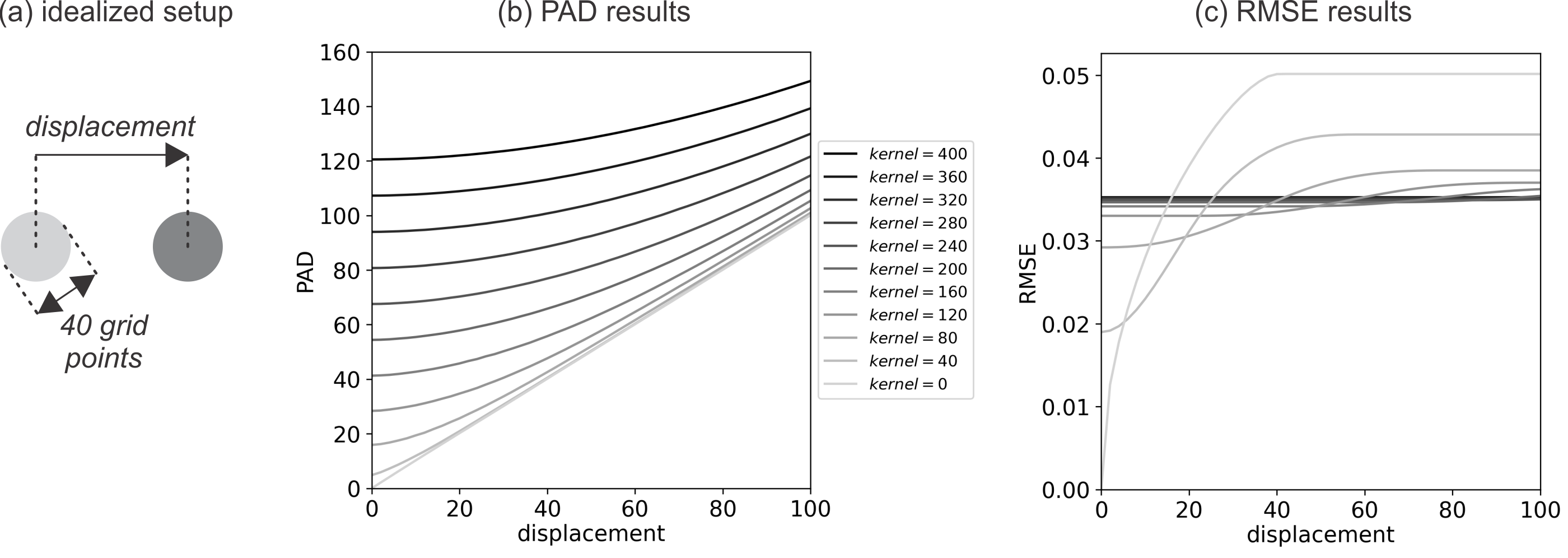}
\caption{(a) The setup of the idealized case with two identical circular events, one in each field, that are displaced. (b) The resulting PAD values with respect to the displacement and the diameter of the circular smoothing kernel (lines with different shades of gray with the corresponding smoothing kernel diameter size denoted in the legend). (c) Same as (b) but for the RMSE.}  
\label{fig:smoothin_analysis_idealized}
\end{figure}

The idealized case is in planar geometry on an equidistant grid. It consists of two identical circular events with a diameter of 40 grid points, one in each field, displaced from 0 to 100 grid points, as shown in Fig~\ref{fig:smoothin_analysis_idealized}a. The grid point values inside each event are set to one and zero everywhere else. One field is then smoothed using a circular averaging kernel (the smoothed value is the sum of all the values inside the kernel divided by the number of grid points inside the kernel) with the diameter of the kernel ranging from 0 (no smoothing) to 400 grid points. 

Fig~\ref{fig:smoothin_analysis_idealized}b shows the resulting PAD values with respect to the displacement of the events and the diameter of the circular smoothing kernel (no distance cutoff is used). The results show that the smoothing always increases the PAD value, although the increase can be very small if the magnitude of the smoothing (size of the smoothing kernel) is small. The increase is also bigger at smaller displacements, indicating that the smoothing has a smaller effect on the PAD value if the displacements are larger. 

For comparison, Fig~\ref{fig:smoothin_analysis_idealized}c shows the results for the  Root-Mean-Square-Error (RMSE) metric. Here, the behavior is more complex. At small displacements, the RMSE value increases if more smoothing is applied, while at larger displacements, the opposite is true. Also, with large smoothing kernels (e.g., kernel diameter larger than 120 grid points), the RMSE becomes almost insensitive to the displacement and kernel size, which does not seem to happen in the case of PAD. Moreover, even at smaller smoothing kernels, the RMSE value does not change further when a certain displacement is reached, meaning the metric becomes insensitive to displacement. This happens once the precipitation from the two events no longer overlaps - for the case with no smoothing, this happens at a displacement of 40 grid points when the two circular events stop overlapping. 

\begin{figure}[btp]
\centering
\includegraphics[width=0.8\textwidth]{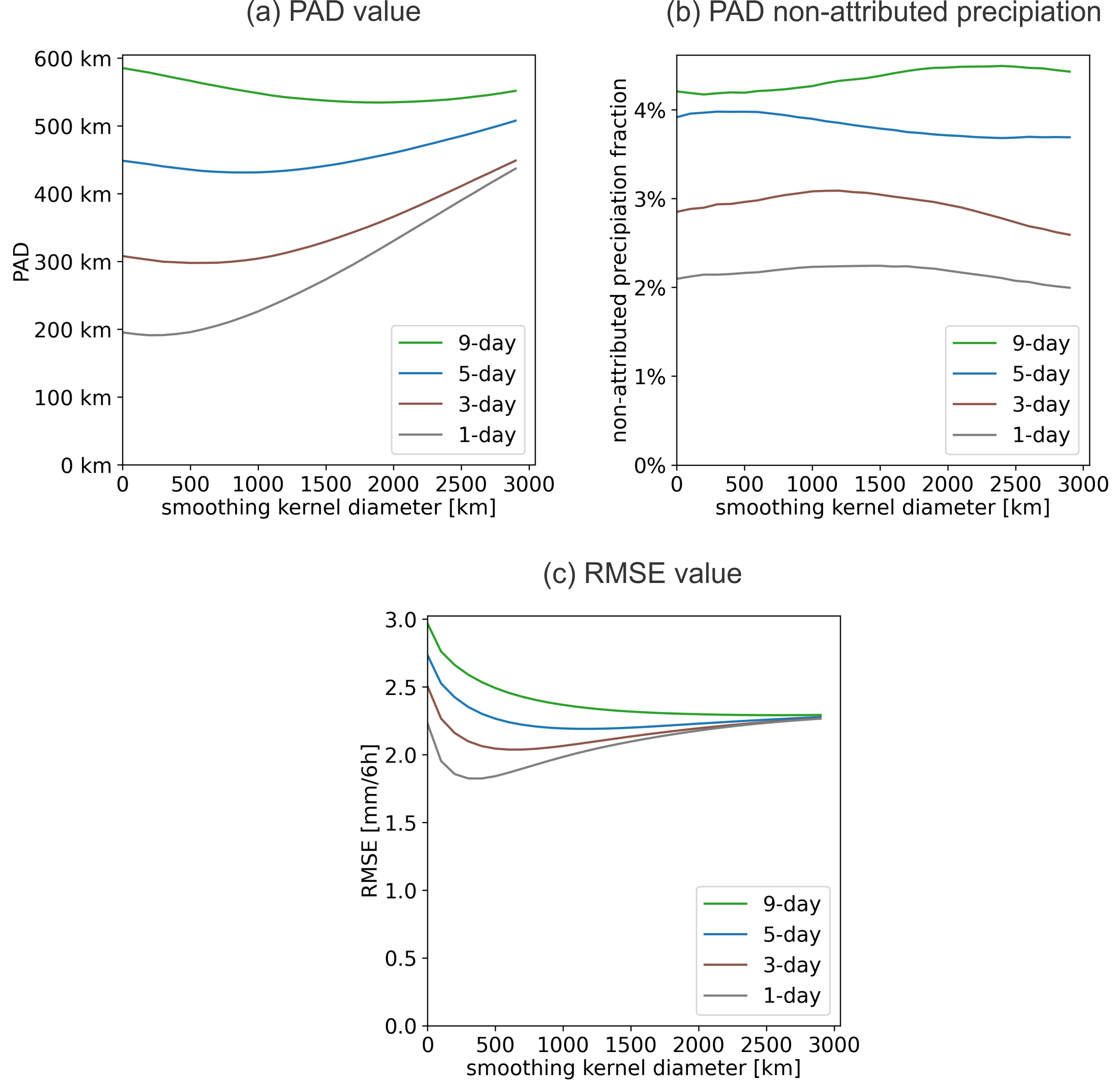}
\caption{The effect of smoothing for the 1-, 3-, 5- and 9-day-ahead IFS forecast of 00-06 UTC precipitation for 11 October 2022. (a) the PAD value, (b) the non-attributed precipitation fraction, and (c) the RMSE value.}  
\label{fig:smoothin_analysis_IFS}
\end{figure}

Figures~\ref{fig:smoothin_analysis_IFS}(a-b) show the influence of smoothing on the PAD value and the amount of non-attributed precipitation for the 1-, 3-, 5- and 9-day-ahead IFS forecast of 00-06 UTC precipitation for 11 October 2022 using a 3000 km distance cutoff (the same case is shown in Fig.5 in the main manuscript and was also used in the hedging analysis shown in Section~\ref{sec:bias_hedging})

As the domain is spherical in this case, with the grid box area not being the same for all points,  the smoothed value at a certain location was calculated as the gridpoint-area-weighted average of precipitation intensities of the points inside a sphere cap area centered at the location. The diameter of the kernel thus represents the along-the-surface diameter of the sphere cap.

Results show that the PAD value tends to increase for the short-term forecasts (e.g., 1- and 3-day-ahead forecasts) if more smoothing is applied, while it can also decrease a bit for longer lead times when the displacements are larger. At the same time, the amount of non-attributed precipitation does not seem to be substantially influenced by the smoothing.  

Fig.\ref{fig:smoothin_analysis_IFS}c shows the results for the RMSE. Here, the RMSE value starts to decrease noticeably at all lead times when some smoothing is applied (although the values for the shorter lead times start to increase eventually), and, eventually, for large smoothing kernels, the RMSE value becomes almost the same for all lead times, meaning that it does not differentiate between the quality of smoothed shorter- and longer-term forecasts, which is not the case for PAD. 

Overall, the analysis of the influence of the forecast's smoothness shows that PAD behaves relatively well (decreases only slightly with smoothing for the longer lead-times) while the behavior of the RMSE is much more problematic, with its value decreasing noticeably due to smoothing in some situations or becoming almost insensitive to the magnitude of displacement in others. 

\clearpage
\section{RMSE-based Timeseries Analysis}

\begin{figure}[b]
\centering
\includegraphics[width=0.9\textwidth]{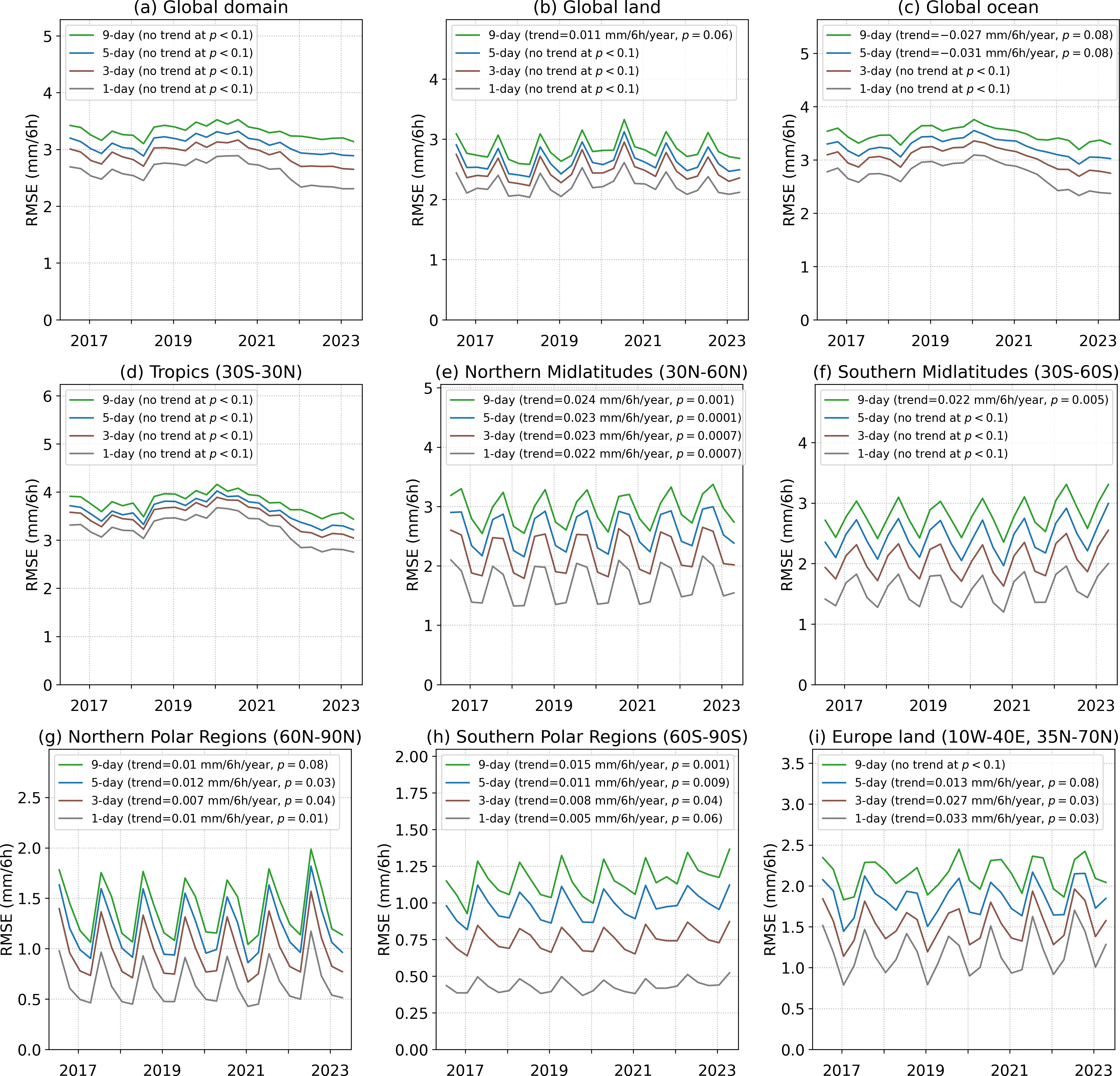}
\caption{Timeseries of mean seasonal RMSE values for different sub-regions. Statistically significant trends are indicated in the legend. A significance level of $p < 0.1$ has been used for all tests.}
\label{fig:IFS_timeseries_seasonal_RMSE}
\end{figure}

Fig.\ref{fig:IFS_timeseries_seasonal_RMSE} is similar to Fig.7 from the main manuscript but shows the time evolution of RMSE metric instead of PAD. As can be observed, there are no statistically significant improvements in forecast performance for any sub-region and any forecast time, with the exception of 9-, and 5-day-ahead forecasts for the Global ocean sub-region. At the same time, there are many examples of deteriorating forecast performance (e.g., Northern Midlatiudes, Northern and Southern Polar Regions, and Europe Land, for most or all forecast times). 

\clearpage
\section{LPAD Analysis}

\begin{figure}[h]
\centering
\includegraphics[width=0.95\textwidth]{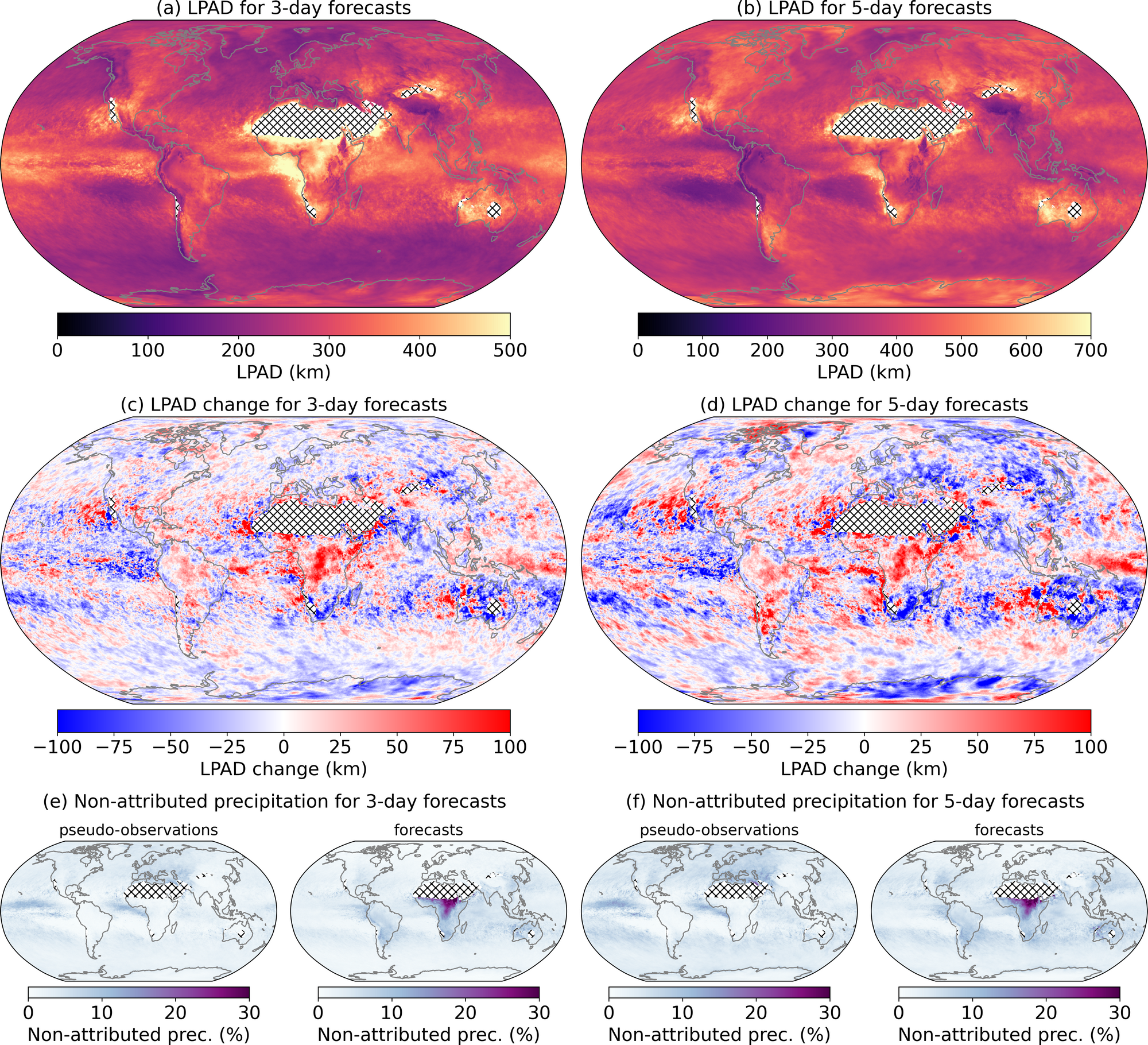}
\caption{Same as Figure 10 from the main manuscript but for 3-, and 5-day forecasts.}
\label{fig:IFS_LPAD_global}
\end{figure}

Fig.\ref{fig:IFS_LPAD_global} is the same as Figure 10 in the the main manuscript, but shows the LPAD results for the global domain for the 3- and 5-day-ahead forecasts.

\begin{figure}[bt]
\centering
\includegraphics[width=0.65\textwidth]{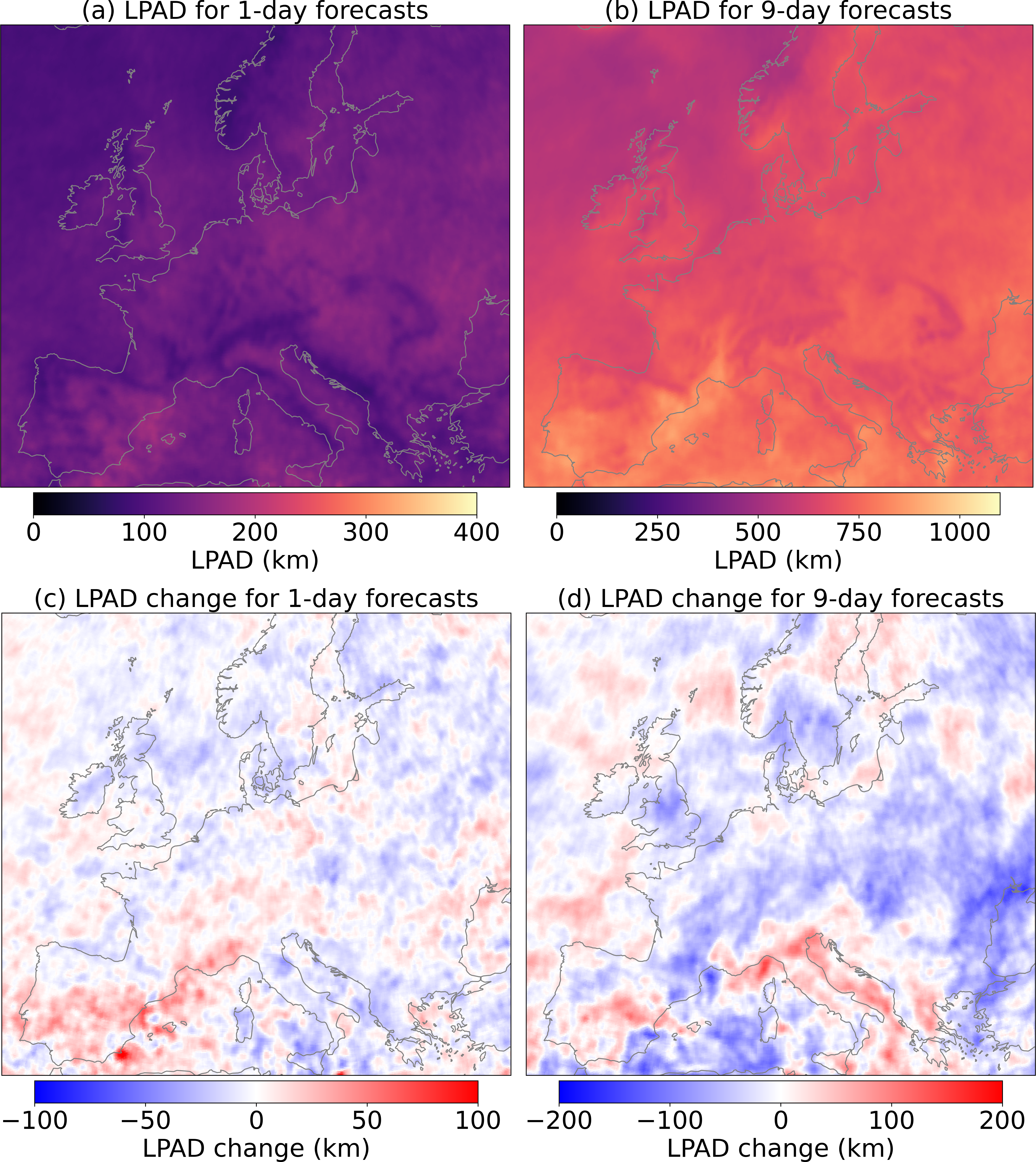}
\caption{Same as Figure 10(a-d) from the main manuscript but with an enlarged view of Europe.}
\label{fig:IFS_LPAD_europe}S
\end{figure}

\begin{figure}[bt]
\centering
\includegraphics[width=0.65\textwidth]{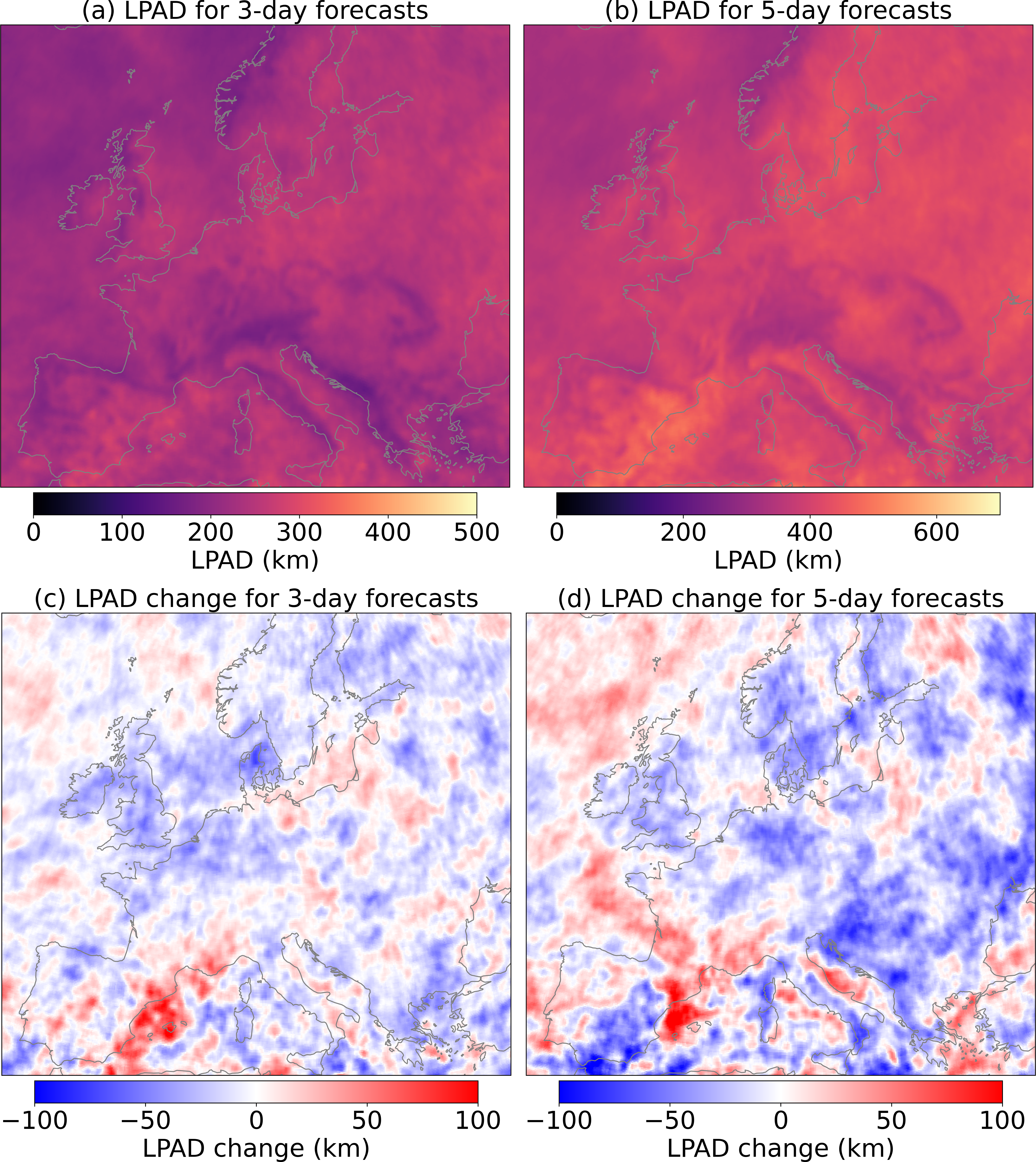}
\caption{Same as Fig.\ref{fig:IFS_LPAD_europe} but for the 3-, and 5-day-ahead forecasts.}
\label{fig:IFS_LPAD_europe2}
\end{figure}

Figs.\ref{fig:IFS_LPAD_europe}-\ref{fig:IFS_LPAD_europe2} show the LPAD results for Europe in more detail. For the 1-day-ahead forecasts (Fig.\ref{fig:IFS_LPAD_europe}a), the largest LPAD values are over the western Mediterranean. The smaller LPAD values are located near the main mountainous regions -- this is likely linked to orographically induced precipitation, which will tend to be less displaced in the short-range forecasts as it is fixed due to the location of the mountains. The 1-day-ahead forecasts improved during the 7-year period in large parts of continental Europe (Fig.\ref{fig:IFS_LPAD_europe}c), with the notable exception of southern France and Spain. 

For the 9-day-ahead forecasts (Fig.\ref{fig:IFS_LPAD_europe}b), the largest LPAD values are again over the Mediterranean and tend to decrease toward the northwest. The effect of the orography is still visible but is less pronounced. The forecasts at this lead time also improved in large parts of continental Europe (Fig.\ref{fig:IFS_LPAD_europe}d), with some exceptions (e.g., Spain, Italy, western and southern Balkans).

\begin{figure}[bt]
\centering
\includegraphics[width=0.8\textwidth]{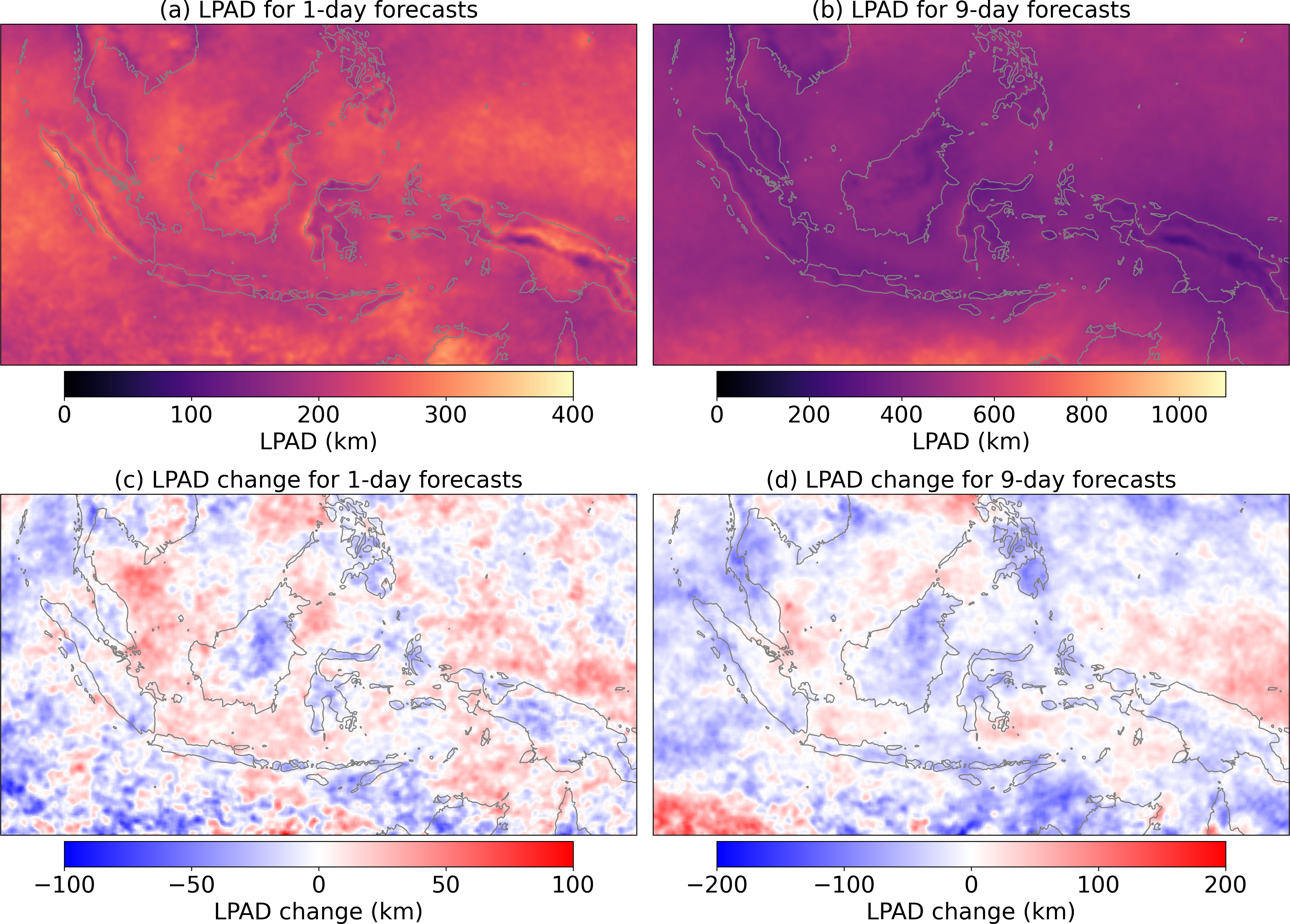}
\caption{Same as Figure 10(a-d) from the main manuscript but with an enlarged view of the Maritime Continent.}
\label{fig:IFS_LPAD_Maritime}
\end{figure}

\begin{figure}[bt]
\centering
\includegraphics[width=0.8\textwidth]{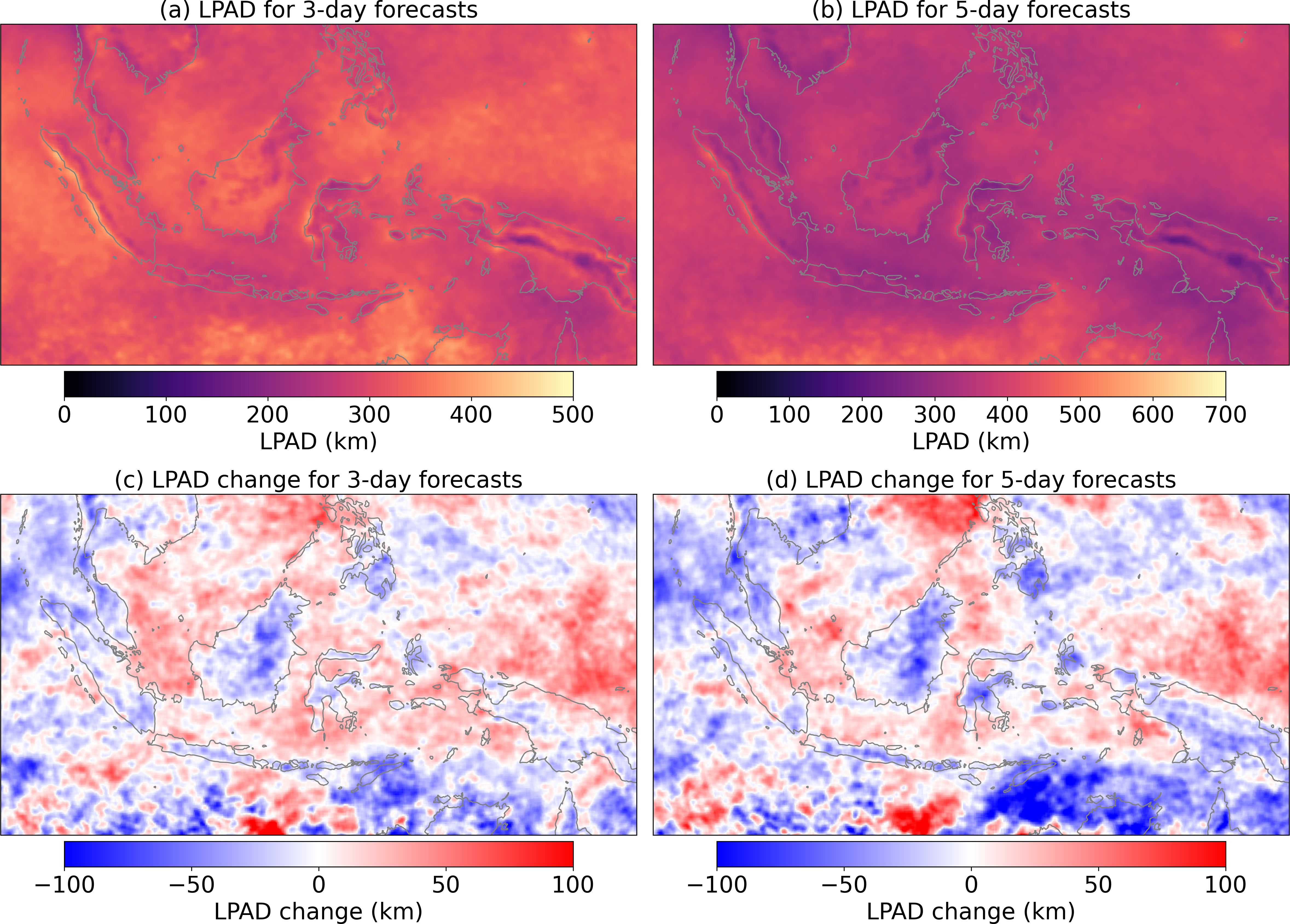}
\caption{Same as Fig.\ref{fig:IFS_LPAD_Maritime} but for the 3-, and 5-day-ahead forecasts.}
\label{fig:IFS_LPAD_Maritime2}
\end{figure}

Figs.\ref{fig:IFS_LPAD_Maritime}-\ref{fig:IFS_LPAD_Maritime2} show the LPAD results for the Maritime Continent in more detail. For the 1-day-ahead forecasts (Fig.\ref{fig:IFS_LPAD_Maritime}a), the largest LPAD values are over or near land regions, especially the  New Guinea island and the east coasts of Sumatra and Sulawesi. Fig.\ref{fig:IFS_LPAD_Maritime}c shows a very high variability of LPAD change, with a mix of red and blue regions, although New Guinea, Sumatra, Sulawesi, and part of Borneo exhibit some improvement. 

For the 9-day-ahead forecasts (Fig.\ref{fig:IFS_LPAD_Maritime}b), the largest LPAD values tend to be over the ocean. Similarly to the results for Europe (Fig.\ref{fig:IFS_LPAD_europe}), there is an orographic effect with smaller LPAD values near mountainous regions (e.g., on the New Guinea island). Remarkably, the 9-day-ahead forecasts improved over almost all land regions of the Maritime Continent (Fig.\ref{fig:IFS_LPAD_Maritime}d).